\documentclass[10pt, twocolumn]{IEEEtran}

\usepackage[T1]{fontenc}% optional T1 font encoding
\ifCLASSINFOpdf
\else
\fi
\usepackage{amsmath}
\newcommand{\req}{\text{req}}
\newcommand{\mb}{\mathbf}
\newcommand{\mc}{\mathcal}
\newcommand{\F}{\mathcal{F}_\text{req}}
\newcommand{\B}{\mathcal{B}}
\newcommand{\K}{\mathcal{K}}
\newcommand{\R}{\mathcal{R}}
\newcommand{\cP}{\mathcal{P}}
\newcommand{\U}{\mathcal{U}}
\newcommand{\E}{\mathbf{E}}
\newcommand{\bt}{\mathbf{t}}
\newcommand{\W}{\mathcal{W}}
\newcommand{\V}{\mathcal{V}}
\newcommand{\br}{\mathbf{r}}
\newcommand{\Z}{\mathbf{Z}}
\newcommand{\vecc}{\text{vec}}
\newcommand{\MDS}{\mathcal{P}_0^\text{MDS}}
\newcommand{\un}{\mathcal{P}_0^\text{FU}}

\usepackage{amsfonts,upgreek,pifont,amssymb,enumerate,color,algorithm,theorem}
\usepackage{algpseudocode}
\usepackage{cite,graphicx,subfigure}
\usepackage{mathtools}
\usepackage{caption,epstopdf}

\DeclarePairedDelimiter\floor{\lfloor}{\rfloor}

\newenvironment{proof}
    {
      \\\emph{Proof.}
    }
    { 

    }

\begin{document}

\title{Joint Fronthaul Multicast and Cooperative Beamforming for Cache-Enabled Cloud-Based Small Cell Networks: An MDS Codes-Aided Approach}

\author{Xiongwei~Wu,~\IEEEmembership{Student~Member,~IEEE},~Qiang~Li,~Victor~C.~M.~Leung,~\IEEEmembership{Fellow,~IEEE},~and~P.~C.~Ching,~\IEEEmembership{Fellow,~IEEE}%
\thanks{
% This paragraph of the first footnote will contain the date on 
% which you submitted your brief for review. It will also contain support 
% information, including sponsor and financial support acknowledgment. For 
% example, ``This work was supported in part by the U.S. Department of 
% Commerce under Grant BS123456.''. 
Part of the work was presented at the IEEE International Conference on Acoustics, Speech, and Signal Processing (ICASSP), Calgory, Canada, April 15 - 20, 2018.}%
\thanks{X. Wu and P. C. Ching are with the Department of Electronic Engineering, Faculty of Engineering, The Chinese University of Hong Kong, Shatin, Hong Kong SAR of China  (e-mail: xwwu@ee.cuhk.edu.hk; pcching@ee.cuhk.edu.hk).}% <-this % stops a space
\thanks{Q. Li is with the School of Information and Communication
Engineering, University of Electronic Science and Technology of China,
Chengdu 611731, China (e-mail: lq@uestc.edu.cn).}% <-this % stops a space
% \thanks{Manuscript received April 19, 2005; revised August 26, 2015.}
\thanks{V. C. M. Leung is with the College of Computer Science and Software Engineering, Shenzhen University, Shenzhen 518060, China, and the Department of Electrical and Computer Engineering, The University of British Columbia, Vancouver, BC, V6T 1Z4 Canada (email: vleung@ieee.org).}
}

\maketitle 
\begin{abstract}
The performance of cloud-based small cell networks (C-SCNs) relies highly on a capacity-limited fronthaul, which degrade quality of service when it is saturated. Coded caching is a promising approach to addressing these challenges, as it provides abundant opportunities for fronthaul multicast and cooperative transmissions. {This paper investigates a cache-enabled C-SCNs, in which small-cell base stations (SBSs) are connected to the central processor via fronthaul, and can prefetch popular contents by applying maximum distance separable (MDS) codes.} To fully capture the benefits of fronthaul multicast and cooperative transmissions, an MDS codes-aided transmission scheme is first proposed. We formulate the problem to minimize the content delivery latency by jointly optimizing fronthaul bandwidth allocation, SBS clustering, and beamforming. To efficiently solve the resulting nonlinear integer programming problem, we propose a penalty-based design by leveraging variational reformulations of binary constraints. To improve the solution of the penalty-based design, a greedy SBS clustering design is also developed. Furthermore, closed-form characterization of the optimal solution is obtained, through which the benefits of MDS codes can be quantified. Simulation results are given to demonstrate the significant benefits of the proposed MDS codes-aided transmission scheme.
% Simulation results demonstrate that the proposed algorithms outperform existing designs under different network settings, and MDS coded design can reduce latency substantially compared with uncoded designs. 
\end{abstract}

% Note that keywords are not normally used for peerreview papers.

% For peer review papers, you can put extra information on the cover
% page as needed:
% \ifCLASSOPTIONpeerreview
% \begin{center} \bfseries EDICS Category: 3-BBND \end{center}
% \fi
%
% For peerreview papers, this IEEEtran command inserts a page break and
% creates the second title. It will be ignored for other modes.
\IEEEpeerreviewmaketitle

\section{Introduction}

\IEEEPARstart{C}{loud-based} {small cell networks (C-SCNs) have been proposed to be a promising architecture for fifth-generation wireless networks (5G), as it offers high data rate and low power consumption \cite{zhang2016fronthauling,hardjawana2016parallel,chen2015cloud}. By connecting multiple small-cell base stations (SBSs) to the central processor (CP) via fronthaul, C-SCNs enable cloud computing and dense deployment of small cells. Hence, these features allow centralized optimization for resource allocation and interference management across multiple SBSs \cite{hardjawana2016parallel}. 
However, using high-speed fronthaul links between the CP and SBSs incurs additional cost, which constitutes a problem given the rapid growth of mobile data traffic.}
% SBSs may fetch a large number of files from a CP during peak hours and 
% The fronthaul link may be saturated, causing long delays in content delivery and degradation of quality of service. Thus, satisfying the requirement of extremely low latency (1 ms or lower) in the 5G system presents a major challenge \cite{shariatmadari2015machine}.

Recently, caching at the physical layer is widely recognized as an effective technique for alleviating traffic burden on the capacity-limited fronthaul. 
% {\red In cache-enabled C-SCNs, the conventional RRHs are replaced by some low-cost small-cell base stations (SBSs) \cite{hu2018joint,quek2017cloud}. These SBSs have the functionalities of caching and signal processing, which can serve users' requests cooperatively under the coordination of CP \cite{hu2018joint,park2016jointJ}.} 
Mobile data traffic is dominated by popular multimedia contents. Many users may request the same popular content, and this practice is referred to as content reuse \cite{golrezaei2012femtocaching,tao2016content}.
Therefore, caching these frequently requested contents at SBSs in off-peak sessions can substantially alleviate fronthaul traffic load. Moreover, delivering these contents without accessing the cloud also assists to reduce delay and power consumption  \cite{tao2016content,peng2017layered,li2018hierarchical}. Hence, in cache-enabled C-SCNs, one needs to study two problems, i.e., content placement and content delivery. Specifically, the content placement phase focuses on how to cache contents so that they can be frequently used over a long period; while the content delivery phase focuses on how to deliver contents so as to satisfy users' requests given the cached status in all SBSs \cite{tao2016content}. 

In general, caching strategies can be classified into two types, i.e., uncoded caching and coded caching. In uncoded caching, each SBS can either fetch the entire files or  fragments of the files from the cloud. By using file splitting, fetching the uncoded fragments can give a good content diversity of cached resources \cite{liao2017coding}, because in practice the cache storage in SBSs is fairly limited compared with the content library in the cloud. Moreover, in this case, file splitting could also increase the cache hit ratio and ensure caching fairness among various kinds of content requests. Given the massive amounts of popular files in the cloud and capacity-limited fronthaul links, coded caching has received increasing research interests recently. Instead of caching entire files or uncoded fragments, the idea of using prefetching coded packets has been demonstrated to have a higher probability to reduce the fronthaul load over uncoded caching \cite{pedersen2019optimizing,liao2017coding}. Specifically, each file is encoded into multiple coded packets by using network code such as maximum distance separable (MDS) codes. Due to the correlation among coded packets, any collection of a certain number of unique coded packets is sufficient to recover the entire file. Therefore, the CP can send a few packets simultaneously to associated SBSs through shared fronthaul bandwidth without requiring a specific delivery order of packets for each SBS. Hence, the coded design provides more opportunities for fronthaul multicast and cooperative transmissions \cite{liao2017coding}.

However, many existing studies on the coded design generally focused on content placement \cite{gabry2016energy,ji2014caching,liao2017optimizing}.  
% To exploit such benefits of coded caching, employing MDS codes to enhance the performance of wireless networks has attracted considerable attention recently.
% Many preliminary studies were developed to design the coded caching strategies in content placement phase.
The study in \cite{bioglio2015optimizing} proposed an MDS coded caching scheme at the wireless edge to minimize backhaul cost without considering fronthaul multicast. The study in \cite{liao2017coding} investigated the advantages of utilizing MDS codes in heterogeneous networks by combining multicast and cooperative content sharing. 
% The studies in \cite{li2018hierarchical,pedersen2019optimizing} investigated the potentials of MDS coded caching in both mobile devices and SBSs to reduce traffic load. 
These studies generally investigated how to store coded packets in terms of optimizing averaged network performance metrics. The physical-layer transmissions, i.e., transmission policy of delivering coded packets to satisfy users' requests, were not well studied. 
Some other works also studied the benefits of coded caching from the perspective information theory \cite{maddah2014fundamental,sengupta2017fog,kakar2017delivery}.
% For instance, by allowing only intra-file coding, the authors in \cite{azimi2018online} studied online strategies for content placement and content delivery to minimize the normalized delivery time 
% , which is widely considered to be effective in high signal-to-noise. 
% Also, the study in \cite{kakar2017delivery} used the NDT to capture the download latency. 
Although content delivery was discussed in these studies, the important problem of beamforming and SBS collaboration in the content delivery phase were not considered. To date, from the perspective of resource allocation and signal processing, the benefits of fronthaul multicast and cooperative transmissions enabled by MDS coded caching have not been well unleashed yet.

To fill the aforementioned research gap, in this paper, under MDS coded caching, we investigate effective content delivery design for cache-enabled C-SCNs. Specifically,  
we study effective strategies for joint fronthaul bandwidth allocation, SBS clustering, and cooperative beamforming, with the goal of achieving low latency. In this MDS codes aided design, users requesting the same content form a multicast group and are served by a cluster of selected SBSs through cooperative beamforming in the edge link. Benefiting from the correlation among MDS coded packets, fronthaul multicast is adopted in the fetching of uncached packets. To balance the fronthaul traffic load and latency among each multicast group, the SBS cluster and fronthaul bandwidth should be judiciously scheduled according to their cached packets and channel state information (CSI). Otherwise, the capacity-limited fronthaul may be saturated and thus causes substantial latency. 
% Benefiting from the centralized signal processing in C-SCNs, we consider that the CP has the knowledge of all users' requested information, CSI and cached packets in the SBSs. 
% In practice, this cooperative transmissions scheme can be scheduled in the cloud through advanced coordinate techniques, such as the coordinated multiple-point (CoMP) technique \cite{checko2015cloud}. 
To this end, we aim to address two fundamental issues: 
{\bf i) by using MDS codes, how fronthaul multicast and cooperative transmissions can improve the efficiency of content delivery in cache-enabled wireless networks; and ii) under MDS coded caching, how efficient optimization algorithms can be developed for better resource allocation.}

% There are some important features that distinguish our work from existing research on content delivery design.
To the best of our knowledge, this is the first work to investigate the potentials of employing MDS codes in physical-layer transmissions by jointly considering fronthaul multicast, beamforming and SBS collaboration. The vast majority of existing content delivery studies focused on beamformers design or SBS collaboration by separate costly unicast transmissions via expensive fronthaul links. For instance, aiming at saving cost and power consumption,
the studies in \cite{tao2016content,peng2017layered}  investigated the content-centric beamforming strategy by  fetching the uncached (entire) files from the CP in unicasting manner. 
The works in \cite{park2016joint,wxw2018cacran} proposed  content delivery schemes to minimize latency by fetching the uncoded fragments via fronthaul unicast. The study in \cite{vu2017joint} also investigated cooperative beamforming under the scenario where the uncached fragments were individually fetched from the CP. Although \cite{liu2013mixed} investigated MDS codes, it considered beamformer design only under two modes, namely, each user was provided service by either all SBSs or one SBS. On the other hand, the authors in \cite{park2017coded} also discussed coded caching in terms of reducing latency. However, the focus was limited to unicast beamformer design in the edge link. Besides, this scenario was also used in \cite{ugur2016cloud}, and only fixed connectivity was considered without using fronthaul bandwidth allocation. In contrast to these content delivery studies, the proposed MDS codes aided design have different features. One critical feature that distinguishes our work from \cite{tao2016content,peng2017layered,park2016joint,wxw2018cacran,vu2017joint} is that given the finite-capacity fronthaul links in C-SCNs, it is essential to investigate using the same radio resource to multicast the uncached contents from the CP. Different from \cite{liu2013mixed,park2017coded,ugur2016cloud}, another feature is that  our work investigates the benefits of joint fronthaul multicast and SBS collaboration. This is because the utilization of fronthaul multicast enables more SBSs to participate in the cooperative transmissions to boost the edge throughputs without incurring extra fronthaul cost. This process helps to improve spectrum efficiency and reduce network latency. Lastly, to ensure latency fairness, fronthaul allocation is performed by adapting to the fronthaul traffic loads. 
These features make our latency-oriented problem more challenging in designing efficient beamformers for cache-enabled C-SCNs.

Moreover, the challenge of the proposed content delivery design is twofold. Firstly, the latency objective in our problem is highly non-convex. Although the studies in \cite{park2016joint,park2017coded} provided a method to decouple the latency objective, it incorporated SBS transmit beamformers only. In the proposed design, the tight coupling among fronthaul bandwidth allocations, SBS beamforming and clustering makes the considered latency objective much more challenging.
% Moreover, both studies adopted semidefinite relaxation (SDR) technique to address the nonconvexity of the data rate expression. This approach is at cost of lifting the dimension of variables and usually incurs high complexity. Also the rank } 
Secondly, owing to SBS clustering, the resultant optimization problem usually falls into mixed-integer nonlinear program (MINLP), which is NP-hard in general.
% Exhaustive search is prohibitive when the problem dimension is large. 
For solving MINLP efficiently, some recent studies have developed sparsity-based methods in wireless system designs \cite{tao2016content,peng2017layered,vu2017joint,hu2017joint,sun2018qoe}, such as the reweighed $l_1/l_2$-minimization approach. 
% {\blue Specifically, this approach essentially utilizes the group sparsity of the multicast beamformers, i.e.,  replacing the SBS clustering variables with the $l_0$-norm of the power of multicast beamformers; and  
% this non-smooth term $l_0$-norm is further approximated by using some smooth functions. 
% {\blue For instance, the reweighed $l_1/l_2$-minimization method was used in \cite{peng2017layered} to solve the network power minimization problem.} 
However, our problem admits a very complicated form, i.e., the multicast data rate function, fractional form of binary variables over continuous ones and so on. Consequently, applying this method may suffer from performance loss due to a relaxation gap between the $l_0$-norm and $l_1/l_2$-norm. 
Another limitation of this method is that the group sparse structure of the resultant beamformers may not be preserved \cite{hu2017joint,peng2017layered}. Hence, existing methods are not readily applied to address the considered problem.
% We will explore new optimization techniques for content delivery design in cache-enabled wireless networks.

The main contributions of this paper are summarized as follows:
\begin{itemize}
  \item 
  To exploit fronthaul multicasting and cooperative beamforming, an MDS codes aided transmission scheme is firstly proposed for cache-enabled C-SCNs.
  With fairness among multicast groups, fronthaul bandwidth allocation, SBS clustering, and beamforming are jointly optimized under physical layer transmission and fronthaul bandwidth constraints. 
  % As a comparison, the uncoded design is also developed by enabling fronthaul unicast. 
  \item A penalty-based design is derived to solve the latency minimization problem efficiently. In particular, to convert the resulting MINLP into tractable forms, we recast the binary constraints into continuous ones by leveraging a variational reformulation of the binary constraints. Building upon the reformulation, { an inexact block coordinate update (BCU) algorithm is developed with convergence guarantee. Moreover, a greedy SBS clustering design is also developed to further improve the solution of the inexact BCU algorithm.}

  \item  
  % Necessary conditions for the optimal solutions are derived for both coded caching design and the uncoded one. 
  % From the conditions, the benefits of MDS codes can be explicitly quantified. 
  Although the MINLP problem is known to be difficult to find the optimal solution, the closed-form characterization of the optimal solution is developed. These closed-form expressions not only serve as a tool to reduce complexity of the proposed algorithm and analyze the impact of some system parameters, but also give insights into the performance gain of employing MDS codes over the alternative concept of uncoded fragments in physical-layer transmissions.
\end{itemize}

% \subsection{Organization and Notations}

The remainder of the paper is organized as follows. Section II describes the system model. Section III shows the problem formulation and analysis. Section IV presents the inexact BCU-SCA design. Closed-form characterization of the optimal solution is shown in Section V. Section VI demonstrates the performance of the proposed designs through numerical experiments. Finally, we conclude the paper in Section VII. 
 
\emph{Notations:} 
% We denote vectors as lower-case bold face letters, and matrices as upper-case bold letters. 
The real part operator and conjugate operator for complex value are  denoted as Re$\{\cdot\}$ and $\{\cdot\}^*$, respectively. 
% Further, operators $\mb A^T$, $\mb A^H$, $\mb A^{-1}$, and $\|\mb A\|_p$ denote the transpose, Hermitian transpose, inverse, and $l_p$ norm of matrix $\mb A$, respectively.
The identity matrix, zero matrix, and 1 matrix are denoted as $\bf I$, $\bf 0$, and $\bf 1$ with appropriate dimensions, respectively. The cardinality of a set is denoted as $|\cdot|$. The operator $\cup$ denotes the union of a collection of sets.
% We use $\mathbb C~(\mathbb{R})$ for the complex (real) space.
The operator $\vecc (\mb A)$ represents the column-wise vectorization of matrix $\mb A$. 

\section{System Model}
\begin{figure}[h]
  \centering
  \includegraphics[scale=0.5]{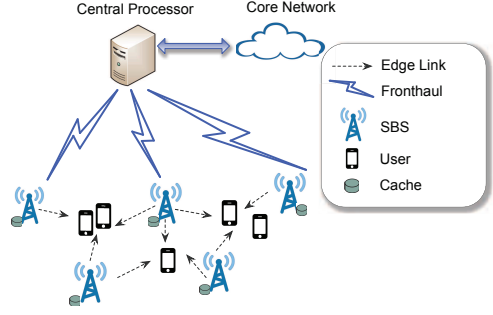}
  \caption{An illustration of downlink cache-enabled C-SCNs.}
  \label{system}
\end{figure}

As shown in Fig. \ref{system}, we consider the downlink transmission of a $B \times K$ cache-enabled C-SCNs, where $K$ users are cooperatively served by a cluster of $B$  densely deployed SBSs on the wireless channels, referred to as the edge link. Each user is equipped with $N$ antennas.
% {\red (revised:where K  users  are cooperatively served by a cluster of B that densely deploys BSs on wireless channels. This cluster is referred to as the edge link. )} 
Besides, each SBS is equipped with $M$ antennas and connects to the CP through the wireless fronthaul link with finite capacity {\cite{hu2018joint}}. 
We assume that the CP has access to the entire library, which stores a total of $F$ files. {For simplicity, each file is assumed to be equal in size $S$ bits.} Let ${\cal{B}} = \{1,\dots, B\}, \mathcal{F} = \{ 1,\dots, F\}$, and $\mathcal{K} = \{ 1,\dots, K\}$ denote the sets of SBSs, files in the library, and users, respectively. We index the files by the order of  popularity, and the probability $p(f)$ of file $f$ being requested by users follows the Zipf distribution
$
p(f) = c{f^{ - \gamma }}, 
$
where $\gamma \geq 0$ is a known skewness parameter; and $c$ is a normalization constant\cite{tao2016content}.
Each SBS has a local cache with a storage of $\mu FS$ bits, where $\mu \in [0, 1]$ is the fractional caching capacity. A cache-aided system usually operates in two phases: the content placement phase and the content delivery phase, which will be described in subsequent subsections. 

\subsection{Content Placement and MDS Coding}

According to the principle of MDS codes, in general, each file is equally split into $n$ fragments and further coded into $m$ packets without information loss, where $m$ can be any integer satisfying $m \geq n $ \cite{liao2017coding,liu2013mixed}. By taking advantage of the dependency among these coded packets, 
any $n$ unique packets are sufficient to recover the original file. Notably, MDS codes can be achieved with only an extremely small redundancy \cite{liu2013mixed}. 

In the cache-enabled C-SCNs, we denote a collection of all the coded packets of file $f$ as $\mc M_f$. Consider that each file $f$ has $m_{f,b}$ unique coded packets cached at each SBS $b$, where $m_{f,b} \leq n$. Notably, when $m_{f,b} = n$, it implies that file $f$ is entirely fetched from SBS $b$. 
Besides, we collect the cached packets of file $f$ in SBS $b$ as the set $\mc M_{f,b}$, satisfying $\mc M_{f,b} \subset \mc M_f$.
% and $0 \leq |\mc M_{f,b}| \leq  |\mc M_f|$
% {\red Overlap of the cached packets in each BS can be alleviated by satisfying $0 \leq m_{f,b} \leq n_f$.}
% Note that for each file $f$, the cached packets in all BSs are different from each other. 
Correspondingly, $n - m_{f,b}$ unique packets are needed to recover file $f$ in SBS $b$. For the exploitation of multicast and reduction of traffic load in the fronthaul link, the CP is required to deliver at least $\max_b ~( n - m_{f,b})$ unique packets, so that each SBS can successfully perform MDS decoding.  Therefore, in the multicast-aware setting, the number of coded packets for file $f$ should satisfy
% \begin{align}
%   {l_f} = \underbrace {\sum\limits_{b \in {\mathcal{B}}} {{m_{f,b}}} }_{{\text{unique packets cached in BSs}}} + \underbrace {{n_f} - \mathop {\min }\limits_{b \in {\mathcal{B}}} {m_{f,b}}}_{{\text{multicast packets delivered via fronthaul}}}
%   \label{MDS}
% \end{align}
\begin{align}
  {m} \geq  \underbrace {\big|\mathop\cup\limits_{b \in {\mathcal{B}}} {{\mc M_{f,b}}} \big|}_{{\text{unique packets cached in BSs}}} + \underbrace {\mathop {\max }\limits_{b \in {\mathcal{B}}} ~ (n -  {m_{f,b}})}_{{\text{multicast packets delivered from the CP}}},
  \label{MDS}
\end{align}
which ensures that the cached packets in the local SBSs and the uncached packets fetched via the fronthaul link are sufficiently different from each other \cite{liao2017coding}. 
% {\red (revised: so that cached packets in local
% BSs are guaranteed, and uncached packets fetched via a fronthaul link are sufficiently different from one another.
% )}

As mentioned previously, MDS codes are helpful to reduce the fronthaul traffic load. To provide greater detail about this, we consider the following simple example:
% To take advantage of the independence among MDS coded packets, it may potentially reduce the fronthaul traffic load.
% The following simple example accounts the properties of MDS codes
file 1 is split into three fragments $\{a_1, a_2, a_3\}$, i.e., $n = 3$, and then coded into packets $\mc M_1 = \{x_i| i = 1,\cdots, 4\}$, i.e., $m = 4$. In addition, there are three SBSs in total, with cached packets $\mc M_{1,1} = \{x_1, x_2\}, \mc M_{1,2} = \{x_2, x_3\}$, and $\mc M_{1,3} = \{x_1,x_3\}$, respectively. 
% {\red (revised: Three BSs have packets $\{x_1, x_2\}, \{x_2, x_3\}$, and $\{x_1,x_3\}$ is cached in each BS.)} 
Consequently, we have $m_{1,b} = 2, \forall b \in \mc B = \{1,2,3\}$.
According to the principle of MDS codes, any three unique packets are sufficient to recover file 1. Subsequently, we can multicast packet $x_4$ to all SBSs via the fronthaul link, and thus file 1 can be recovered at each SBS. Meanwhile, for the uncoded design, each SBS stores fragments directly. For instance, fragments $\{a_1, a_2\}, \{a_2, a_3\}$, and $\{a_1,a_3\}$ are cached in each SBS. As a result, all fragments $a_1, a_2$, and $a_3$ are transferred in the fronthaul link for file reconstruction in each SBS, and this process increases traffic load.

In practice, in order to reduce complexity and implementation cost, the CP usually focuses on designing cache allocation matrix $\mb Q =[q_{f,b}]$, where each element $q_{f,b} \triangleq m_{f,b}/n $ indicates the fraction of file $f$ that is randomly cached in SBS $b$ \cite{liao2017coding,liu2013mixed,bioglio2015optimizing}. In general, cache allocation $\mb Q$ needs to be designed by taking into account the file popularity, the finite storage in each SBS and so on. Notably, the file popularity distribution should vary much slower than the channel conditions of wireless links. For instance, a film may remain popular for several days. 
Hence, the cached resource is considered to be static, when we design a transmission scheme to satisfy users' requests in a much shorter period. In our content delivery design, we consider that cache allocation matrix $\mb Q$ is fixed by using certain popular strategies \cite{park2016jointJ}, and elaborate on signal processing and resource allocation in the content delivery phase. Similar modeling assumption has been used in previous study (see, e.g., \cite{peng2017layered,sun2018qoe,nguyen2019collaborative}).

\subsection{Content Delivery}
% {\blue As mentioned above}, we focus on content delivery design with  knowledge of cached packets in each SBS, i.e., $\forall m_{f,b}$. 
% being a known priori. 
% For the exploitation of the benefits of MDS codes and cooperative transmissions, fronthaul bandwidth allocation, content assignment in fronthaul links and multigroup multicast beamforming are jointly optimized. 
% This strategy achieves low latency.
For ease of implementation, the content delivery phase is assumed to operate in a transmission-interval fashion. At the beginning of each transmission interval, each user $k$ requests a certain file $f_k$.
The set $\mathcal{F}_{\text{req}} \subset \mc F$ denotes the indexes of requested files from all users, with cardinality $F_{\req} = |\mathcal{F}_{\req}|$. 
Then, users requesting the same file are formed a multicast group. Specifically, the users
in multicast group $f$ are denoted as $\mathcal{G}_f$, and they are only requesting file $f$. 
For this reason, we have $\mathcal{G}_{f_1} \cap \mathcal{G}_{f_2} =\emptyset$, for $\forall f_1 \neq f_2$. 
We define a SBS clustering matrix $\E = [ e_{f,b}] \in \{ 0,1 \}^{F_{\req} \times B}$, where each element $e_{f,b} = 1 $ indicates that SBS $b$ is selected to serve the multicast group $f$, otherwise 0. 
% Thus, the cluster of BSs for multicast group $f$ is defined as the set $\{ \mathcal{B} |~ e_{f,b} = 1, \forall b \in \mathcal{B}\}$.
Besides, each file is allowed to be cooperatively served by a cluster of SBSs, i.e.,
\begin{align} 
   \textstyle \sum_{b\in \B} e_{f,b} \geq 1, ~ \forall f \in \F.   
\end{align}  
For the edge link, 
the signal transmitted from SBS $b$ is 
    \begin{align}
       {{\mathbf{x}}_b} = \textstyle \sum_{f \in {\mathcal{F}_{{\text{req}}}}} {{{\mathbf{v}}_{f,b}}{x_f}}, 
    \end{align}
where $\mathbf{v}_{f,b} \in \mathbb{C}^M$ denotes the multicast beamformer for file $f$ at SBS $b$, and the signal $x_f \in \mathbb{C}$ independently encodes the file $f$ for multicast group $f$, with distribution $ x_f \sim \mathcal{N}(\mathbf{0},\mathbf{I})$.
% Denote the network-wide beamformers that precode fragment $(f,l)$ from all BSs as  
% ${{\mathbf{V}}_{f,l}} =[ {{\mathbf{V}}_{f,l}^1;{\mathbf{V}}_{f,l}^2; \cdots ;{\mathbf{V}}_{f,l}^J} ]$. 
%And $d$ is the dimension of the coded signals which satisfies $d  \leq \min\{M, N \}$    
% Each fragment is independently coded, and user can reconstruct the file by receiving all  fragments \cite{khreishah2016joint}. 
% Note that if fragment $(f,l)$ is not served by $i$-th BS, the corresponding beamformer $\mathbf{V}_{f,l}^i$ should be $\mathbf{0}$. 
%Based on the assignment variable $e_{f_k,l}^i$,  we have 
%    \begin{align}
%        \left( {1 - e_{f_k,l}^i} \right){{\mathbf{V}}_i}^{f_k,l} = {\mathbf{0}},\label{ZF}
%    \end{align}
%which shows that if the $i$-th BS does not transfer such fragment $(f_k,l )$ to $k$-th user, the corresponding beamformer is zero matrix.
Accordingly, the ${{\mathbf{v}}_f} = \left[ {{\mathbf{v}}_{f,1}^H,{\mathbf{v}}_{f,2}^H,\cdots,{\mathbf{v}}_{f,B}^H} \right]^H$ denotes the aggregate beamformer from all SBSs serving the transmission of file $f$. In particular, if SBS $b$ does not serve the transmission of file $f$, the corresponding transmit beamformer $\mathbf{v}_{f,b}$ should be $\mathbf{0}$. Subsequently, we have
\begin{align}
(1 - e_{f,b})\mathbf{v}_{f,b} = \mathbf{0}. \label{Coupling}
\end{align} 
User $k$ applies the linear combiner $\mathbf{u}_k \in \mathbb{C}^{N}$ to mitigate inter-group interference, which yields the received signal
    % \begin{align}
% {{{y}}_k} &= \underbrace {{\mathbf{u}}_k^H\left( {\textstyle\sum_{b \in {\mathcal{B}}}^{} {{{\mathbf{H}}_{kb}}{{\mathbf{v}}_{{f_k},b}}} } \right){x_{{f_k}}}}_{{\text{Desired signal}}} \notag\\
% &+ \underbrace {{\mathbf{u}}_k^H\textstyle\sum_{f \in {\mathcal{F}_{{\text{req}}}}\backslash \{ {f_k}\} } {\left( {\textstyle\sum_{b \in {\mathcal{B}}}^{} {{{\mathbf{H}}_{kb}}{{\mathbf{v}}_{f,b}}} } \right){x_f}} }_{{\text{Interference}}} + {\mathbf{u}}_k^H{{\mathbf{z}}_k} \label{Signal},
%     \end{align}
%%% Signal version
%     \begin{align}
% {{{y}}_k} &= \underbrace {{\mathbf{u}}_k^H\left( {\sum\limits_{b \in {\mathcal{B}}}^{} {{{\mathbf{H}}_{kb}}{{\mathbf{v}}_{{f_k},b}}} } \right){x_{{f_k}}}}_{{\text{Desired signal}}} + \underbrace {{\mathbf{u}}_k^H\sum\limits_{f \in {\mathcal{F}_{{\text{req}}}}\backslash \{ {f_k}\} } {\left( {\sum\limits_{b \in {\mathcal{B}}}^{} {{{\mathbf{H}}_{kb}}{{\mathbf{v}}_{f,b}}} } \right){x_f}} }_{{\text{Interference}}} + {\mathbf{u}}_k^H{{\mathbf{z}}_k} \label{Signal},
%     \end{align}
    \begin{align}
{{{y}}_k} &= \underbrace {{\mathbf{u}}_k^H \mb H_k \mb v_{f_k}{x_{{f_k}}}}_{{\text{desired signal}}} + \underbrace {{\mathbf{u}}_k^H\sum\limits_{f \in {\mathcal{F}_{{\text{req}}}}\backslash \{ {f_k}\} } {\mb H_k\mb v_f{x_f}} }_{{\text{inter-group interference}}} + {\mathbf{u}}_k^H{{\mathbf{z}}_k} \label{Signal},
    \end{align}
where aggregate channel matrix ${{\mathbf{H}}_k} = \left[ {{{\mathbf{H}}_{k1}},{{\mathbf{H}}_{k2}}, \cdots ,{{\mathbf{H}}_{kB}}} \right]$, and ${{{\mathbf{H}}_{kb}}} \in \mathbb{C}^{N \times M}$ denotes the channel matrix between SBS $b$ and user $k$, for any $b$; and $\mathbf{z}_k$ denotes the additive complex Gaussian noise with distribution $\mathbf{z_k} \sim \mathcal{CN}(\mathbf{0},\sigma_k^2\mathbf{I})$. For notational simplicity, we define the signal variable $D_k = {\chi_{k,1}}(\mb u_k, \mb v_{f_k})$ and the covariance as $J_k = {\chi_{k,2}} (\mb u_k, \V)$, where
    % \begin{align}
    %     D_k & =  {\mathbf{u}}_k^H{{\mathbf{H}}_k}{{\mathbf{v}}_{{f_k}}},\\
    %     J_k & =  {\mathbf{u}}_k^H\left( {\sum\limits_{f \in {\mathcal{F}_{{\text{req}}}}\backslash \{ {f_k}\} }^{} {{{\mathbf{H}}_k}{{\mathbf{v}}_f}{\mathbf{v}}_f^H{\mathbf{H}}_k^H} } \right){{\mathbf{u}}_k} + \sigma _k^2,
    %  \end{align}
    \begin{align}
        {\chi_{k,1}}(\mb u_k, \mb v_{f_k}) &= {\mathbf{u}}_k^H{{\mathbf{H}}_k}{{\mathbf{v}}_{{f_k}}},\\
        {\chi_{k,2}} (\mb u_k, \V) &= {\mathbf{u}}_k^H\left( {\textstyle\sum_{f \in {\mathcal{F}_{{\text{req}}}}\backslash \{ {f_k}\} }^{} {{{\mathbf{H}}_k}{{\mathbf{v}}_f}{\mathbf{v}}_f^H{\mathbf{H}}_k^H} } \right){{\mathbf{u}}_k} + \sigma _k^2.
     \end{align}     
Assuming that the receiver regards the interference as noise, the achievable data rate for multicast group $f$ is given by
\begin{align}
  {R_f}  = \mathop {\min }\limits_{k \in {{\mathcal{G}}_f}} B_0\phi \left( D_k, J_k \right),
\end{align}
where function $\phi ({D_k},{J_k}) = \log \left( {1 + {{\left| {{D_k}} \right|}^2}/{J_k}} \right)$; and $B_0$ is the bandwidth.
  % Denote the set of data rates of all users as $ \big\{\mathcal{R}|R_k, k \in \mathcal{K}\big\}$.

% {\blue When the requested files are not entirely cached in the local BSs, the missing packets are fetched from the CP by fronthaul multicasting.} 
Regarding fronthaul transmission, to ensure that the CP can serve multiple SBSs simultaneously, we adopt frequency division multiplexing to schedule the multicast service in the fronthaul link.
 % , instead of time division multiplexing access (TDMA) used in \cite{hu2017joint}. 
Notably, when the requested files are not entirely cached in the selected SBSs, the missing packets should be fetched from the CP via fronthaul.\footnote{We consider the case where each SBS needs to collect the missing packets of the requested file so as to recover the entire file, similar to \cite{liao2017coding,li2018hierarchical,vu2017joint}. In this case, multiple SBSs are clustered on file level and use cooperative beamforming to deliver the same information to the dedicated users, which can boost the edge throughputs. If different SBSs directly transmit distinct fragments or packets to users,  all SBSs are clustered on fragment/packet level. In other words, all SBSs
need to dynamically form different clusters in order to complete the transmission of each fragment/packet. This process may cause excessive signaling overhead and  synchronization cost in practice.}
For ease of discussion, we consider that the total capacity of the fronthaul link is limited by $C_F$ bps. In addition, the fronthaul radio frequency is different from the edge link;
hence no interference is observed between the fronthaul and edge links. 
As a result, a fraction of the fronthaul bandwidth is allocated to serve the cluster of SBSs for multicast group $f$, 
\begin{align}
  R_{f}^{\text{fh}} = t_f C_F, ~\textstyle\sum_{f \in \mathcal{F}_{\text{req}}} t_f = 1,~0 \leq t_f \leq 1, 
\end{align}  
where $t_f$ is the ratio of the fronthaul bandwidth allocated to deliver file $f$; and $R_{f}^{\text{fh}}$ denotes the rate for multicast group $f$ in the fronthaul link. We define vector $\bt = [t_f] \in \mathbb{R}^{F_{\req}}$. 
 % where the cluster of BSs serving for multicast group $f$ is defined as the set $\{ \mathcal{B}_f |~ e_{f,b} = 1, \forall b \in \mathcal{B}\}$. 
% The condition \eqref{eq:SF} can also be rewritten as 
% \begin{align}
%   S_F^f = \mathop {\max }\limits_{b \in \mathcal{B}} {\text{ }}{e_{f,b}}\left( {1 - \frac{{{m_{f,b}}}}{{{n_{f}}}}} \right)S ~\text{bits}
% \end{align}
% From \eqref{eq:SF}, the coupling of the content assignment and 
Note that some packets have already been cached in the local SBSs, and thus the traffic load for file $f$ to be transmitted from the CP to SBS $b$ should be 
% \begin{align}
%   s_{f,b} = e_{f,b}m_{f,b}',  
% \end{align}
$
  s_{f,b} = e_{f,b}m_{f,b}',  
$
where 
$
 m_{f,b}' = \left(  1- q_{f,b} \right) S  ~\text{bits}. 
$ Define the traffic load matrix $\mb S = [s_{f,b}] \in \mathbb{R}^{F_{\req}\times B}$. 
Based on the principle of MDS codes, the fronthaul load for multicast group $f$ is given by
% \begin{align}
%   S_F^f = \mathop {\max }\limits_{b \in \mathcal{B}} {\text{ }}  e_{f,b}m_{f,b}', 
%   % {e_{f,b}}\left( {1 - \frac{{{m_{f,b}}}}{{{n_{f}}}}} \right)S ~\text{bits} 
%   \label{eq:SF}
% \end{align}
\begin{align}
  s_f = \mathop {\max }\limits_{b \in \mathcal{B}} {\text{ }}  e_{f,b}m_{f,b}'.
  % {e_{f,b}}\left( {1 - \frac{{{m_{f,b}}}}{{{n_{f}}}}} \right)S ~\text{bits} 
  \label{eq:SF}
\end{align}
% at least. Note that when $S_F^f = 0$ bit, we have the $t_f = 0$ and  $R_{f}^{\text{fh}} = 0$. 
% Recalling equation \eqref{Coupling}, we can obtain the coupling of content assignment and beamforming, i.e., the traffic load assignment $\{S_F^f\}$ and the beamformer $\{\mb v_{f,b}\}$ should be jointly designed.
The multicast fronthaul load vector is further defined as $\mb s = [s_f] \in \mathbb{R}^{F_{\req}}$.   
Note that when $s_f = 0$ bit, we have the $t_f = 0$ and  $R_{f}^{\text{fh}} = 0$. 
By recalling 
% {\red (As shown in)} 
\eqref{Coupling},  a proper SBS clustering matrix not only balances the fronthaul traffic load, but also increases spatial diversity in the edge links.

% coupling of content assignment and beamforming, i.e., the traffic load assignment $\{s_f\}$ and the beamformer $\{\mb v_{f,b}\}$ should be jointly designed.  

\subsection{Network Latency Model}
In this work, we focus on the latency caused by content delivery through the fronthaul link and edge link, and ignore the latency due to geometry propagation, requests queuing, and other factors. The latency in the network is further defined as the number of symbols or channel uses needed to accomplish the requested files transmission \cite{park2016joint,park2017coded,koh2017cloud}. We assume that the
messages delivery is half-duplex [15]; thus, the system operates in
a serial mode. 
% \footnote{The proposed design can also be applied in a pipeline mode, where the latency is evaluated as $\max \{T_E, T_F\}$ \cite{kakar2017delivery}. The proposed algorithms still work.}. 
In other words, the CP first communicates with SBSs
and is then followed by edge transmissions. For balancing traffic load in the network, min-max fairness is considered in our model. With SBS cooperation and multicast beamforming, edge latency is evaluated as
    \begin{align}
         {T_E} = \frac{S}{{{{\min }_{f \in {\mathcal{F}_{{\req}}}}}{R_f}}},  \label{eq:TE}
     \end{align}
where the minimization is over all multicast groups.
Regarding the multicast transmission via fronthaul, when file $f$ is fully cached in the selected BS cluster, i.e., $s_{f} =0$ and $R_f^{\text{fh}} =0$, no fronthaul latency is detected in multicast group $f$. {To balance fronthaul traffic load, fronthaul latency is evaluated as}
    \begin{align}
         {T_F} = \mathop {\max }\limits_{f \in {\mathcal{F}_{{\text{req}}}}, R_{f}^{\text{fh}} \neq 0} ~{{s_f}}/{ R_{f}^{\text{fh}} }, \label{eq:TF}
     \end{align}
under the condition of \eqref{MDS}.
Denote $\U=\{ \mathbf{u}_k, \forall k \in \mathcal{K}\}$, $\V = \{\mathbf{v}_{f}, \forall f \in \mathcal{F}_{\req}\}$.
Therefore, by adopting fronthaul multicast and cooperative beamforming, the total latency of the network is evaluated as
    % \begin{align}
    %     T_{{\text{total}}}^{\text{MDS}} \left(\U, \V, \mathbf{E}, \mathbf{t}\right) &=   \alpha_E \frac{S}{\mathop{\min}\limits_{f \in \F} \mathop {\min }\limits_{k \in {{\mathcal{G}}_f}} B_0\phi \left( D_k, J_k \right)}\notag\\ 
    %     &+ \alpha_F \mathop{\max }\limits_{f \in {\mathcal{F}_{{\text{req}}}},b \in \mathcal{B}} ~ \frac{e_{f,b} m_{f,b}'}{t_{f}C_F + \tau _0}, \label{eq:Ttot} 
    % \end{align}
    % Single version
    \begin{align}
        &T_{{\text{total}}}^{\text{FM}} \left(\U, \V, \mathbf{E}, \mathbf{t}\right) \notag= \\&\alpha_E \frac{S}{\mathop{\min}\limits_{f \in \F} \mathop {\min }\limits_{k \in {{\mathcal{G}}_f}} B_0\phi \left( D_k, J_k \right)} 
        + \alpha_F \mathop{\max }\limits_{f \in {\mathcal{F}_{{\text{req}}}},b \in \mathcal{B}} ~ \frac{e_{f,b} m_{f,b}'}{t_{f}C_F + \tau _0}, \label{eq:Ttot}
    \end{align}
where $\tau_0 >0$ is a small constant that is used for preventing zero denominator, when a null load is present in the fronthaul link for certain multicast groups. Weights $\alpha_E > 0$ and $\alpha_F > 0$ balance the importance of the latency in the edge link and fronthaul link, respectively. 
Similar performance metrics have been used in previous studies \cite{li2015distributed,koh2017cloud,park2017coded}.
% ding2017network 

% To minimize the network latency, the content assignment and beamformers are optimized in CP. The content assignment is characterized by determining which fragment $(f,l)$ should be transferred through BSs. If the fragment $(f, l)$ is not cached in the $i$-th BS, it should be accessed via the fronthaul link. The assignment is completed by setting the binary variables $e_{f,l}^i =1 $ if fragment $(f,l)$ is transferred by the $i$-th BS, otherwise $0$. Define the variable $d_{f,l}^i$ as $
%         d_{f,l}^i = (1 - c_{f,l}^i)e_{f,l}^i$
% to indicate that fragment $(f,l)$ is accessed via fronthaul link at the $i$-th BS when $d_{f,l}^i = 1$. We can observe that when $c_{f,l}^i = 1$, fragment $(f,l)$ can directly be sent from the $i$-th BS without fronthaul latency. Thus, the size of the content transferred via the fronthanl link to the $i$-th BS is given by
% $
% {S_F^i}                  = {\sum_{f \in {\mathcal{F}_{{\text{req}}}}}^{} {\sum_{l \in \mathcal{L}} {d_{f,l}^i{S_{  L}}} } },
% $
% %    \begin{align}
% %                {S_F^i} = {\sum\limits_{f \in {\mathcal{F}_{{\text{req}}}}}^{} {\sum\limits_{l \in \mathcal{L}} {d_{f,l}^i{S_{  L}}} } },
% %    \end{align}
% where each fragment is assumed to have the same size of $S_L$ bits.

% We only focus on content delivery design, with the knowledge of cached files in all BSs, i.e.,  $m_{f,b}$ being known a priori.  
% The delivery phase is defined as follows. 

% clue: the initial problem -> analysis -> penalty-based reformulation. 

\section{Problem formulation and analysis}
% In this section, we propose an MDS codes aided scheme for content delivery, which aims at minimizing the total latency in the network. To fully capture the benefits of MDS codes and BS cooperation, we optimize beamforming (i.e., $\U$, $\V$), SBS clustering (i.e., $\mathbf{E}$), and fronthaul bandwidth allocation (i.e., $\mathbf{t}$) simultaneously.

\subsection{Problem Formulation for MDS Coded Design}

% This incorporates the joint design of the transmit beamformers, the receiver combiners, as well as the assignment of the transfering contents of  all small BSs. 
In the considered cache-enabled C-SCNs, all CSI, the cached contents in the local SBSs, and the requested files are known at the CP. Hence the optimization problem is stated as

% For notation simplicity,  define set $\mathcal{E}= \big\{ {{e_{f,b}}}| f \in {\mathcal{F}_{{\text{req}}}},l \in \mathcal{L},i \in \mathcal{J} \big\}$, set $\mathcal{V}= \big\{ {{\mathbf{V}_{f,l}^i}}| f \in {\mathcal{F}_{{\text{req}}}},l \in \mathcal{L},i \in \mathcal{J} \big\}$, and set $\mathcal{U}= \big\{ {{\mathbf{U}_k}}| f \in k \in \mathcal{K} \big\}$, thus the problem is stated as
\begin{subequations}
  \label{MinT}
  \begin{align}
    \MDS:  \mathop {\min }\limits_{\U,\V,\mathbf{E},\mathbf{t}} &~T_{{\text{total}}}^{\text{FM}} \left(\U, \V, \mathbf{E}, \mathbf{t}\right) \label{P_a}\\
    \text{s.t.}   &\textstyle\sum_{f \in {\mathcal{F}_{{\text{req}}}}} {\left\| {{{\mathbf{v}}_{f,b}}} \right\|_2^2}  \leq {P_b},b \in {\mathcal{B}}, \label{P_b}\\
    &\left\| {{{\mathbf{u}}_k}} \right\|_2 = 1, k \in \mathcal{K} \label{P_c},\\
    &\left( {1 - {e_{f,b}}} \right){{\mathbf{v}}_{f,b}} = {\mathbf{0}},{\text{  }}\forall b \in \mathcal{B}, f \in {\mathcal{F}_{{\text{req}}}}, \label{P_d}\\
    &{e_{f,b}} \in \left\{ {0,1} \right\},{\text{    }}\forall b \in {\mathcal{B}}, f \in {\mathcal{F}_{{\text{req}}}}, \label{P_e}\\
    & \textstyle \sum_{b \in \mathcal{B}} e_{f,b} \geq 1, \forall f, \label{P_f}\\
    &\textstyle \sum_{f \in {\mathcal{F}_{{\text{req}}}}} {{t_f}}  = 1, 0 \leq t_f \leq 1,  {f \in {\mathcal{F}_{{\text{req}}}}}, \label{P_g}
    \end{align} 
    \end{subequations}
where constraint \eqref{P_b} shows that the peak transmit power for the $b$-th SBS is limited by $P_b$. The receive combiner $\mathbf{u}_k$ has unit power, which is utilized for the mitigation of interference among different multicast groups. 
Constraint \eqref{P_d} reflects the relation between beamformer design and SBS clustering. 
% It should be noted when the $b$-th BS is not assigned to serve multicast group $f$, the corresponding beamformer $\mb{v}_{f,b}$ should be $\mathbf{0}$. 
Constraint \eqref{P_f} is used to exploit BS cooperation and prevent a standstill service. The bandwidth allocation for all multicast groups is provided by fronthaul constraint \eqref{P_g}.

\subsection{Problem Reformulation and Analysis}
As can be seen that the latency objective \eqref{P_a} in problem $\MDS$ is combinatorial in nature, which incorporates the binary variables, i.e., $\mb E$, and continuous ones, i.e., $\mc U, \mc V, \mb t$. Obtaining the optimal solution is very difficult. This is because even if we use exhaustive search on $2^{BF_\req}$ possible BS clustering matrix $\mb E$, the resulting subproblems are still highly nonconvex. 
% Thus, these observations make our problem more complex than the preliminary studies on latency (see, e.g., \cite{park2016joint,park2017coded}). 

Therefore, to make the objective \eqref{P_a} in ${\MDS}$ more tractable, we introduce some slack variables $t_E, t_F$, and thus problem  $\MDS$ is equivalently expressed as 
% \left\{ \mb{u}_k, \mb{v}_{f,b}, e_{f,b},\mb{t}_f, t_E, t_F\right\}
\begin{subequations}
  \label{P_01}
  \begin{align}
    \mathop {\min }\limits_{\mc{U}, \mc{V}, \mb{E}, \mb{t}, t_E, t_F} &\alpha_E t_E + \alpha_F t_F\\
    \text{s.t.   }  ~~&  t_E \geq \frac{S}{B_0\phi \left( D_k, J_k \right)}, ~~ \forall k, \label{P_01b}\\
            & t_F \geq  \frac{e_{f,b} m_{f,b}'}{t_{f}C_F + \tau _0}, ~~\forall f, b, \label{P_01c}\\
            & \eqref{P_b}- \eqref{P_g},
    \end{align} 
\end{subequations}
by using the epigraph reformulation. For the nonconvex constraint \eqref{P_d}, it can be written as
\begin{align}
  \|\mb{v}_{f,b}\|_2^2 \leq e_{f,b} P_b, ~~ \forall f, b   \label{P_01d}. 
\end{align} 
Furthermore, constraint \eqref{P_01b} can be decoupled by introducing another slack variables $\{r_k\}$, which yields 
\begin{align}
  \log( t_E ) + \log (r_k) \geq \log(S), ~\forall k, \label{P_01e}\\
  r_k \leq B_0\phi \left( D_k, J_k \right), ~\forall k.  \label{P_01f}
\end{align} 
We denote set $\mb{r} =\{ r_k, \forall k \in \mc{K}\}$ and set $\Theta = \{\U, \mc{V}, \mb{E}, \mb{t}, \mb{r},t_E, t_F\}$. Finally, the original problem $\MDS$ is equivalent transformed as 
\begin{subequations}
  \label{P_1}
  \begin{align}
    \mc{P}_1: &\mathop {\min }\limits_{\Theta} ~~ \alpha_E t_E + \alpha_F t_F\\
    \text{s.t.   }  ~~ & \eqref{P_b}, \eqref{P_c}, \eqref{P_e}-\eqref{P_g}, \eqref{P_01c}, \eqref{P_01d}-\eqref{P_01f}.
    \end{align} 
\end{subequations}

By now, problem $\mc{P}_1$ is still the MINLP. The difficulties arise from the involvement of binary variables $\{e_{f,b}\}$  and the noncovexity of the rate function in constraint \eqref{P_01f}. A ready approach for tackling the difficulty of binary variables is to replace each $e_{f,b}$ by $\|\|\mb v_{f,b}\|^2\|_0$ and thus convert the MINLP to the sparse optimization problem \cite{peng2017layered,tao2016content}. Nevertheless, such a replacement will make our problem more complex than itself due to these coupling constraints  \eqref{P_f}, \eqref{P_01c}, \eqref{P_01d}. Besides, when solving such a complex problem, it may suffer from performance loss due to the relaxation gap between term $\|\|\mb v_{f,b}\|^2\|_0$ and approximated term, e.g., reweighed $l_1/l_2$-norm in \cite{peng2017layered}. Regarding the data rate function incorporated in our problem, it was linearized by using semidefinite relaxation (SDR) technique in previous studies \cite{park2016joint,park2017coded}. However, this method usually incurs a high complexity because it is done at a cost of lifting the dimensions of beamformer $\mb v_f$. Moreover, the probability of finding a low rank solution is fairly small as the number of users grows large \cite{karipidis2008quality}. This also probably leads to the solution far from optimal. Considering the challenges identified above, developing a working algorithm is necessary but obviously non-trivial.

\section{Proposed Inexact BCU-SCA Design}
% In this section, instead of applying SDR, we develop the inexact BCU-SCA design to reduce complexity and obtain a near-optimal result. In particular, to make the problem tractable, we adopt another variational reformulation of the binary constraints and a quadratic bound for the rate constraints.
% Similarly, the monotonic convergence is also guaranteed in this penalty-based design.
% In this section, different from applying SDR technique in previous design, 
% we develop an inexact BCU-SCA design to reduce the complexity and obtain the near-optimal result. Specifically, the binary constraint in problem $\cP_1$ are reformulated by a variational method. Due to the nice structure of the resulting problem $\cP_2$, variables are further divided into two blocks, and updated by  BCD method. For one block variable, we again have the closed-form expression to update, while the other block can be inexactly updated by solving a convex quadratic programming directly, instead of the SDR technique. Similarly, the proposed algorithm is also guaranteed with the property of the monotonic convergence. Finally, we will analyze the complexity of the proposed algorithm.

% {\red In this section, instead of applying commonly used SDR technique, we develop the inexact BCU-SCA design with low complexity and obtain a near-optimal result. In particular, to make the problem tractable, we adopt a variational reformulation of the binary constraints \eqref{P_e} and a quadratic bound for the rate constraints \eqref{P_01f}. 
% }
{ In this section, we develop two efficient algorithms to handle problem $\cP_1$ by applying some of the variational reformulations of binary constraint \eqref{P_e}, the successive convex approximation (SCA) technique, and the inexact BCU method. }

\subsection{A Variational Method}
To combat the discontinuity in problem $\mc P_1$, we first introduce  a variational reformulation of the binary constraints. \\[+0.1cm]
{\bf \emph{Lemma 1:}} \emph{We define set $\Omega  = \big\{ ( \E, \Z)  \big| (2 \vecc(\E) - \mb 1)^T (2 \vecc(\Z) - \mb 1) = BF_{\req}. ~\|2\Z - \mb 1\|_F^2 \leq BF_{\req}, \mb 0 \leq \E \leq \mb 1, \Z \in \mathbb{C}^{B \times F_{\req}}\big\}$. If $(\E, \Z) \in \Omega$, then we have $\E \in \{0,1\}^{B \times F_{\req}}$ and $\Z = \E$. }

\vspace*{0.1cm}

% \begin{Lemma}
%  We define set $\Omega  = \big\{ ( \E, \Z)  \big| (2 \vecc(\E) - \mb 1)^T (2 \vecc(\Z) - \mb 1) = BF_{\req}. ~\|2\Z - \mb 1\|_F^2 \leq BF_{\req}, \mb 0 \leq \E \leq \mb 1, \Z \in \mathbb{C}^{B \times F_{\req}}\big\}$. If $(\E, \Z) \in \Omega$, then we have $\E \in \{0,1\}^{B \times F_{\req}}$ and $\Z = \E$. 
% \end{Lemma}
Lemma 1 is a generalized form of Lemma 1 in \cite{yuan2016binary}, which can be proved by Cauchy-Schwarz inequality. 
% is omitted due to space limitations. 
Accordingly, by using an auxiliary matrix $\Z$, problem $\cP_1$ can be rewritten as 
\begin{subequations}
   \label{RR}
  \begin{align}
    \mathop {\min }\limits_{\Theta, \Z} ~~ &\alpha_E t_E + \alpha_F t_F\\
    \text{s.t.   }  ~~ &0 \leq e_{f,b} \leq 1, ~\forall b, f, \label{RR_b}\\
    &  (2 \vecc(\E) - \mb 1)^T (2 \vecc(\Z) - \mb 1) = BF_{\req}, \label{RR_c}\\ 
    & \|2\Z - \mb 1\|_F^2 \leq BF_{\req}, \label{RR_d}\\
    & \eqref{P_b}, \eqref{P_c}, \eqref{P_f}, \eqref{P_g}, \eqref{P_01c}, \eqref{P_01d}-\eqref{P_01f} \label{RR_e}.
    \end{align}\end{subequations}
Although problem \eqref{RR} turns to be continuous, the resulting equilibrium constraint \eqref{RR_c} still makes the problem difficult to handle. 
Hence, we  construct the following  penalty function
% to the previous section
\begin{align}
  h_1(\E, \Z) = BF_{\req} - (2 \vecc(\E) - \mb 1)^T (2 \vecc(\Z) - \mb 1),
\end{align}
which is always non-negative, for any feasible $(\E, \Z)$ satisfying constraints \eqref{RR_b} and \eqref{RR_d}. Taking advantage of this property, the proposed design can be reformulated as 
\begin{subequations}
  \begin{align}
    \R_1: \mathop {\min }\limits_{\Theta, \Z} ~~ &\alpha_E t_E + \alpha_F t_F + \lambda h_1(\E,\Z)\\
    \text{s.t.   }  ~~
    & \eqref{RR_b}, \eqref{RR_d}, \eqref{RR_e} \label{RR_f},
    \end{align}
\end{subequations}
where $\lambda > 0$ serves as a penalty parameter. { Notably, an optimal solution $(\E^*, \Z^*)$ meets the condition $h_1(\E^*, \Z^*) = 0$; for any other $(\E, \Z)$ feasible to constraints \eqref{RR_b} and \eqref{RR_d}, we have $h_1(\E^*, \Z^*) > 0$.} Hence, we tackle problem $\R_1$ by iteratively increasing $\lambda$ to penalize the violation of the equilibrium constraint \eqref{RR_c}, resulting in binary values of SBS clustering matrix $\E$.
% The details for the algorithm are presented in the next subsection.

% Accordingly, one can always enforce $e_{f,b} - e_{f,b}^2 = 0$, by gradually increasing the penalty parameter $\lambda$, resulting in a binary solution. In the following subsections, we will derive an efficient algorithm to solve problem $\mc{R}_0$.
%-----------------------------------------------------------------------
%  \subsection{Inexact BCU-SCA Design}
%-----------------------------------------------------------------------
\subsection{Inexact BCU-SCA Design}
% It is can be observe that the penalty function $h_1(\mb E, \mb Z)$ is a bi-linear term.
In this subsection, we decompose the sophisticated problem $\mc R_1$ into two subproblems: the first one is used to update variables $\{ \mc V, \mb E\}$ without any increase in dimensions of beamformers
while the other one is used to update other variables $\{\mc U, \mb t, \mb Z\}$ with closed-form expressions.

First, we solve the original problem $\cP_1$ by addressing problem $\R_1$ with a fixed penalty parameter $\lambda$. 
% To be explicit, the variables are divided into two block variables, i.e., $\{ \U, \bt, \Z\}$ and $\{ \V, \E\}$. 
When the block $\{ \U, \bt, \Z\}$ is fixed, problem $\R_1$ can be simplified as the following subproblem 
\begin{subequations}
  \begin{align}
    \R_2(\overline \U, \overline  \bt, \overline \Z): \notag
    \mathop {\min }\limits_{\V, \E, t_E, t_F, \br} &~ \alpha_E t_E + \alpha_F t_F + \lambda h_1(\E,\overline  \Z)\\
    \text{s.t.   }~&  t_F \geq  \frac{e_{f,b} m_{f,b}'}{\overline t_{f}C_F + \tau _0}, ~~\forall f, b,  \label{R4_b}\\
    &r_k \leq B_0\phi \left(  D_k', J_k' |\overline \U\right), ~\forall k, \label{R4_c}\\
    & \eqref{P_b},\eqref{P_f}, \eqref{P_01d}, \eqref{P_01e},    
    \eqref{RR_b}, \label{R4_d}
    \end{align}
\end{subequations}
where $\{ \overline \U, \overline \bt, \overline \Z\}$ denotes the solution in the last iteration, $ D_k' = \chi_{k,1} (\overline {\mb u}_k, \mb v_{f_k} )$, and $J_k' = \chi_{k,2} (\overline {\mb u}_k, \V)$.
% \begin{align}
%           D_k'  & =  \overline { \mathbf{u}}_k^H{{\mathbf{H}}_k}{{\mathbf{v}}_{{f_k}}},\\
%          J_k'  & =  \overline {\mathbf{u}}_k^H\left( {\sum\limits_{f \in {\mathcal{F}_{{\text{req}}}}\backslash \{ {f_k}\} }^{} {{{\mathbf{H}}_k}{{\mathbf{v}}_f}{\mathbf{v}}_f^H{\mathbf{H}}_k^H} } \right){\overline {\mathbf{u}}_k} + \sigma _k^2.
% \end{align}
The main challenge of the subproblem $\R_2$ is the non-convexity in constraint \eqref{R4_c}. 

Subsequently, we seek an inner approximation of the constraint \eqref{R4_c}. % By applying Theorem 1 in \cite{tam2016successive}, we have the following quadratic minorant of the rate function $\phi ( D_k' ,  J_k' |\overline \U)$, 
To get rid of the relaxation gap introduced by the SDR technique,
 we consider the following quadratic minorant function of the rate function $\phi ( D_k' ,  J_k' |\overline \U)$, 
\begin{align}
  \widetilde{\phi}_k (\V|\overline \U, \overline \V) = q_0 + 2 \text{Re} \{ D_k'  q_1^* \} - ( | D_k' |^2 + J_k' )q_2, \label{quad_lb}
\end{align}
for $\forall k \in \K$, at any local point $\{\overline \U,  \overline \V\}$, where constants 
% \begin{align}
%   q_0 & = {\phi} (\overline D_k, \overline J_k) -  |\overline D_k|^2/ \overline J_k,\\
%   q_1 & = \overline D_k/\overline J_k,\\
%   q_2 & = 1/\overline J_k - 1/(\overline J_k + |\overline D_k|^2) \geq 0,
% \end{align}
% \begin{align}
%   q_0  = {\phi} (\overline D_k, \overline J_k) -  |\overline D_k|^2/ \overline J_k,~~
%   q_1  = \overline D_k/\overline J_k,~~
%   q_2  = 1/\overline J_k - 1/(\overline J_k + |\overline D_k|^2) \geq 0,
% \end{align}
  $
  q_0  = {\phi} (\overline D_k, \overline J_k) -  |\overline D_k|^2/ \overline J_k,~
  q_1  = \overline D_k/\overline J_k,~
  q_2  = 1/\overline J_k - 1/(\overline J_k + |\overline D_k|^2) \geq 0,
  $
$\overline D_k = \chi_{k,1}(\overline{\mb u}_{f_k}, \overline \V)$, and  $\overline J_k = \chi_{k,2}(\overline{\mb u}_{f_k}, \overline \V)$.
% \begin{align}
%           \overline D_k  & =  \overline { \mathbf{u}}_k^H{{\mathbf{H}}_k}{\overline{\mathbf{v}}_{{f_k}}},\\
%          \overline J_k  & =  \overline {\mathbf{u}}_k^H\left( {\sum\limits_{f \in {\mathcal{F}_{{\text{req}}}}\backslash \{ {f_k}\} }^{} {{{\mathbf{H}}_k}{\overline{\mathbf{v}}_f}\overline{\mathbf{v}}_f^H{\mathbf{H}}_k^H} } \right){\overline {\mathbf{u}}_k} + \sigma _k^2.
% \end{align}

% \begin{align}
%   \phi ( \overline D_k ,  \overline J_k |\overline \U)= \widetilde{\phi}_k (\overline \V|\overline \U, \overline \V). 
% \end{align}

{Equation \eqref{quad_lb} is provided by the theorem 1 in \cite{tam2016successive}.} It holds true that $\phi ( D_k' ,  J_k' |\overline \U)$ is always lower bounded by the quadratic minorant $\widetilde{\phi}_k (\V|\overline \U, \overline \V) $, and the equality holds at the local point $\{\overline \U,  \overline \V\}$, i.e., $
  \phi ( \overline D_k ,  \overline J_k |\overline \U)= \widetilde{\phi}_k (\overline \V|\overline \U, \overline \V). $ Accordingly, problem $\R_2$ can be inner approximated by the following convex program
\begin{subequations}
  \begin{align}
    \widetilde \R_2(\overline \U, \overline \V, \overline  \bt, \overline \Z): 
    \mathop {\min }\limits_{\V, \E, t_E, t_F, \br}  &\alpha_E t_E + \alpha_F t_F + \lambda h_1(\E,\overline  \Z)\\
    \text{s.t.}~~   
    & 
    r_k \leq B_0  \widetilde{\phi}_k (\V|\overline \U, \overline \V),  ~\forall k, \\
    & \eqref{R4_b}, \eqref{R4_d},
    \end{align}
\end{subequations}
which can be efficiently solved by interior point methods using standard solvers, such as CVX \cite{grant2008cvx}. Now we turn to the other subproblem of updating the block $\{\U, \bt, \Z\}$. 
% {\red revised: ( 
% The other subproblem in the updating of block $\{\U, \bt, \Z\}$ is considered.)}
In particular, when $\{\V, \E\}$ are fixed, the resulting subproblem is given by 
\begin{subequations}
  \label{Relax_5}
  \begin{align}
    \mc{R}_3 (\overline{\V}, \overline \E): \mathop {\min }\limits_{ \mb{U}, \mb{t}, \Z, t_E, t_F, \mb{r}} &\alpha_E t_E + \alpha_F t_F  + \lambda h_1(\overline \E,\Z) \\
    \text{s.t.}   ~~& {t_F} \geq \frac{{{{\overline e }_{f,b}}m_{f,b}'}}{{{t_{f}}{C_F} + {\tau _0}}}, \forall f,b, \label{nR3_b}\\
    &r_k \leq B_0\phi \left(  D_k', J_k'|\overline \V \right), \forall k, \label{R5_c}\\
    & \eqref{P_c}, \eqref{P_g}, \eqref{P_01e}, \eqref{RR_d},
    \end{align}
\end{subequations}
where the point $\{\overline \V, \overline \E\}$ denotes the solution in the last iteration, 
$ D_k' = \chi_{k,1} (\mb u_k, \overline {\mb v}_{f_k} )$, and $J_k' = \chi_{k,2} (\mb u_k, \overline \V)$.
% \begin{align}
%            D_k'  & =   { \mathbf{u}}_k^H{{\mathbf{H}}_k}{\overline{\mathbf{v}}_{{f_k}}},\\
%           J_k'  & =   {\mathbf{u}}_k^H\left( {\sum\limits_{f \in {\mathcal{F}_{{\text{req}}}}\backslash \{ {f_k}\} }^{} {{{\mathbf{H}}_k}{\overline{\mathbf{v}}_f}\overline{\mathbf{v}}_f^H{\mathbf{H}}_k^H} } \right){ {\mathbf{u}}_k} + \sigma _k^2.
% \end{align}
We claim that subproblem $\R_3$ has a closed-form solution.\\[+0.1cm]
{\bf \emph{Proposition 1}:} \emph{ Define the multicast traffic load vector $\overline {\mb s} = [\overline s_f] $, where the f-th element $\overline s_f =  \max_b ~\overline e_{f,b} m'_{f,b},  \forall {f \in \F} $.  
Consider the case where constant $\tau_0$ approaches 0 and $\overline {\mb s} \neq  \mb 0$. An optimal solution $(\U^*, \bt ^*, \Z^*)$ to problem $\mc{R}_3(\overline{\V}, \overline{\mb{E}})$ is given as
\begin{align}
  {\mathbf{u}}_k^* &= {{ (\overline{\mb J}_k)^{ - 1}{{\mathbf{H}}_k}{{\overline {\mathbf{v}} }_{{f_k}}}}}\big/{{\big\| { (\overline{\mb J}_k)^{ - 1}{{\mathbf{H}}_k}{{\overline {\mathbf{v}} }_{{f_k}}}}\big\|_2}}, ~\forall k,  \label{OP:u2}\\
  % t_f^* &= 
  % \frac{\max\limits_b ~\overline e_{f,b} m'_{f,b}}{ \|\mb s\|_2} \label{OP:t2}\\
   t_f^* &= 
  \frac{\max\limits_b ~\overline e_{f,b} m'_{f,b}}{ \|\overline {\mb s} \|_1}, ~ \forall f \in \F, \label{OP:t}\\
  {{\mathbf{Z}}^*} &=  
  \begin{cases}
  \text{any feasible value}, ~~ \text{if each} ~ \overline e_{f,b} = \frac{1}{2}, \forall f, b, \\
    \frac{\sqrt{B F_{\req}} (2 \overline \E - \mb 1)}{2\|2 \overline {\mathbf{E}} - \mb 1 \|_F} + \frac{1}{2}, ~~~~\text{otherwise},
  \end{cases}\label{OP:z2}
\end{align}
where 
% \begin{align}
  $ 
  \overline{\mb J}_k  =   {\textstyle\sum_{f \in {\mathcal{F}_{{\text{req}}}}\backslash \{ {f_k}\} }^{} {{{\mathbf{H}}_k}{\overline{\mathbf{v}}_f}\overline{\mathbf{v}}_f^H{\mathbf{H}}_k^H} }  + \sigma _k^2 \mb {I}
  $, for $\forall k$. 
% \end{align}
}
\begin{proof}
See Appendix A.
\end{proof}
% It is worth pointing out that there exists decoupling property among variables $\U$, $\bt$, and $\Z$ in subproblem $\R_3$. To Take advantage of this nice property and make the proposed algorithm more efficient, the block $\{\U, \bt, \Z \}$ is updated in parallel in the proposed design.
Clearly, these closed-form solutions can greatly reduce the computational complexity of the proposed algorithm. Moreover, it is worth pointing out that the three variables $(\U^*, \bt ^*, \Z^*)$ are decoupled in \eqref{OP:u2}-\eqref{OP:z2}. Hence, they can be updated in parallel. Finally, we adopt the inexact BCU method to update both block variables, and the whole procedure for this design is shown in Algorithm 1. Although we can start with any feasible point, the penalty parameter $\lambda >0 $ is generally initialized as a small value for a proper starting point.
% \begin{algorithm}[h]
% \caption{Inexact BCU-SCA Design for Problem \eqref{MinT}}
% \begin{algorithmic}[1]\label{AL1}
% \State {\bf Initialize} $i =0$ , $\U^{(0)}, \V^{(0)}, \Z^{(0)}, \bt^{(0)}$, $\lambda > 0, \eta >1, I$
% \State {\bf Repeat}
% \State  ~~~~Solve problem  $\widetilde{\mc{R}}_2({\U}^{(i)}, \V^{(i)}, \bt^{(i)}, \Z^{(i)})$ to obtain an
% \Statex ~~~~optimal solution $\V^{(i + 1)}, \E^{(i+1)} $ by CVX
% \State ~~~~Solve problem $\mc{R}_3 ({\V}^{(i +1 )}, {\mb{E}}^{(i +1)})$ to obtain an optimal 
% \Statex ~~~~solution $\bt^{(i + 1)}, {\U}^{(i+1)}, {\mb{Z}}^{(i+1)}$ by \eqref{OP:t}, \eqref{OP:u2}, and \eqref{OP:z2}
% \State ~~~~Update $\lambda \leftarrow \lambda \times \eta$ every $I$ iterations

% \State  ~~~~$i \leftarrow i+1$  
% \State {\bf Until} {some stopping criterion is satisfied}
% % \State {\bf Until} {Some stopping criterion is satisfied}
% \State {\bf Output}  $\U^{*}, \V^{*}, \E^{*}, \bt^{*}$
% \end{algorithmic}
% \end{algorithm}
%%%%%%Single version
%%%%%%%%%%%%%%%%%%%%%%%%%%%%%%
\begin{algorithm}[h]
\caption{Inexact BCU-SCA Design for Problem \eqref{MinT}}
\begin{algorithmic}[1]\label{AL1}
\State {\bf Initialize} $i =0$ , $\U^{(0)}, \V^{(0)}, \Z^{(0)}, \bt^{(0)}$, $\lambda > 0, \eta >1, I$
\State {\bf Repeat}
\State ~~~~~Solve problem  $\widetilde{\mc{R}}_2({\U}^{(i)}, \V^{(i)}, \bt^{(i)}, \Z^{(i)})$ to obtain an optimal solution $\V^{(i + 1)}, \E^{(i+1)} $
\State ~~~~~Solve problem $\mc{R}_3 ({\V}^{(i +1 )}, {\mb{E}}^{(i +1)})$ to obtain an optimal solution $\bt^{(i + 1)}, {\U}^{(i+1)}, {\mb{Z}}^{(i+1)}$

by \eqref{OP:u2}, \eqref{OP:t}, and \eqref{OP:z2}
\State ~~~~~Update $\lambda \leftarrow \lambda \times \eta$ every $I$ iterations

\State  ~~~~~$i \leftarrow i+1$  
\State {\bf Until} {some stopping criterion is satisfied}
% \State {\bf Until} {Some stopping criterion is satisfied}
\State {\bf Output}  $\U^{*}, \V^{*}, \E^{*}, \bt^{*}$
\end{algorithmic}
\end{algorithm}
{Obviously, any feasible solution for $\widetilde{\mc{R}}_2$ is also feasible for problem $\R_1$, but the reverse claim does not necessarily hold. Therefore, the optimal value of $\widetilde \R_2$ normally serves as an upper bound of that of problem $\R_1$. For any fixed penalty parameter $\lambda > 0$, i.e., $\eta = 1$, Algorithm 1 generates a sequence  $ \{  \U^{(i)}, \V^{(i)}, \E^{(i)}, \bt^{(i)}, \Z^{(i)}\}$, which gives a monotonically non-increasing objective value of problem $\R_1$. When $\eta > 1$, $\lambda$ will be lifted gradually to enforce the satisfaction of $h_1(\E, \Z) = 0$. Therefore, the sequence $\{\E^{(i)}\}$ generated by Algorithm 1 eventually converges to binary values. }
% We will illustrate the convergence behavior of both two proposed designs in greater details. 
Regarding complexity, it is dominated by solving the convex problem $\widetilde \R_2$ in step 3. Accordingly, it can be solved in polynomial time with regard to  the size of the problem, i.e., the dimension of involved variables $\{ \V, \E, t_E, t_F, \mb r\}$, which is in the order $\mc{O}((M + 1)BF_\req + 2 + K)$.
% {\red If we adopt the conventional SDR approach to tackle the nonconvexity of the rate constraints \cite{park2016jointJ}, i.e., $\mb W_f = \mb V_f \mb V_f^H, \forall f \in \F$, the dimension of involved variables would be approximately squared, which is in the order $\mc{O}((M^2B + 1)BF_\req + 2 + K)$. }As a result, the inexact BCU-SCA design can be implemented with a low computational complexity as well as memory cost.
% {\bf Greedy-SCA:} Given that the optimal value of the proposed design $\MDS$ requires an exhaustive search, we obtain the near-optimal results by a greedy approach. Specifically, starting with the SBS clustering matrix provided by the inexact BCU-SCA design, we iteratively remove one redundant access link in each step, and thus the problem is re-optimized. This procedure is terminated until the result cannot be improved. Accordingly, the complexity will be $\mc O(B^2F_{\req}^2)$ of the complexity of the inexact BCU-SCA design.

\subsection{Greedy SBS clustering Design}
Although the proposed Inexact BCU-SCA design can generally guarantee that each element of beamformer $\mb v_{f,b}$ converges to 0, when $e_{f,b} = 0$, the global optimal solution for $\MDS$ is still hard to obtain.
To further improve the performance of the proposed design, we develop a greedy SBS clustering (GBSC) design, which is built upon the solution generated by Algorithm 1. In cache-enabled wireless networks, the transmit beamformers at SBSs usually admit a group sparse structure \cite{peng2017layered,hu2017joint}, which indicates a sparse connectivity in the edge network. Besides, a group sparse structure of the transmit beamformers is likely to bring along benefits such as reduction in power consumption and signaling overhead. Motivated by these facts,
the crucial idea of GBSC design is as follows: starting with the SBS clustering matrix $\E^* = [e_{f,b}^*]$ provided by the inexact BCU-SCA design, we iteratively remove one redundant edge link in each step, and thus the problem is re-optimized. This procedure is terminated until the objective value cannot be improved; the detailed implementation of GBSC is given in Algorithm 2.
% \begin{algorithm}[h]
% \caption{Greedy SBS clustering Design for Problem (14)}
% \begin{algorithmic}[1]\label{AL1}
% \State {\bf Initialize} $\mc E_0 = \{(f,b)| e_{f,b}^* = 1\},~T_{\min} = + \infty, r = 0$
% \State {\bf Iterate}
% \State ~~~~$\overline {\mc E} \leftarrow \mc E_r$
% \State ~~~~{\bf If} select any $(\bar f, \bar b) \in \overline{\mc E}$, and $\E(\overline{\mc E} \backslash \{(\bar f, \bar b)\})$ is infeasible to problem $\R_1$, {\bf go to step 6}
% \State ~~~~~Set $\overline{\mc E}  \leftarrow \overline{\mc E} \backslash \{(\bar f, \bar b)\}$, and obtain objective value $T(\bar f, \bar b)$ of problem $\R_1$ by inexact BCU-SCA $~~~~$
% with fixed $\E = \E(\mc E_r \backslash \{(\bar f, \bar b)\})$ and $\Z = \E$. {\bf Go to step 4}
% \State ~~~~$(f_r,b_r) = \arg\min_{(f,b) \in \mc E_r} T(f,b)$
% \State ~~~~{\bf If} $T_{\min} > T(f_r, b_r) $
% \State ~~~~~~~~$T_{\min} \leftarrow T(f_r, b_r), \mc E_r \leftarrow \mc E_r \backslash \{(f_r,b_r)\}$, and $~r \leftarrow r+1$  
% \State ~~~~{\bf Else}  terminate
% % \State {\bf Until} {Some stopping criterion is satisfied}
% \State {\bf Output}  $\E(\mc E_r)$
% \end{algorithmic}
% \end{algorithm}
%%% Single Collumn
\begin{algorithm}[h]
\caption{Greedy SBS clustering Design for Problem (14)}
\begin{algorithmic}[1]\label{AL1}
\State {\bf Initialize} $\mc E_0 = \{(f,b)| e_{f,b}^* = 1\},~T_{\min} = + \infty, r = 0$
\State {\bf Iterate}
\State ~~~~$\overline {\mc E} \leftarrow \mc E_r$
\State ~~~~{\bf If} select any $(\bar f, \bar b) \in \overline{\mc E}$, and $\E(\overline{\mc E} \backslash \{(\bar f, \bar b)\})$ is infeasible
to problem $\R_1$, {\bf go to step 6}
\State ~~~~Set $\overline{\mc E}  \leftarrow \overline{\mc E} \backslash \{(\bar f, \bar b)\}$, and obtain objective value
$T(\bar f, \bar b)$ of problem $\R_1$ by inexact $~~~~~~$
$~~~~~$BCU-SCA
with  fixed $\E = \E(\mc E_r \backslash \{(\bar f, \bar b)\})$ and $\Z = \E$. {\bf Go to step 4}
\State ~~~~$(f_r,b_r) = \arg\min_{(f,b) \in \mc E_r} T(f,b)$
\State ~~~~{\bf If} $T_{\min} > T(f_r, b_r) $
\State ~~~~~~~~$T_{\min} \leftarrow T(f_r, b_r), \mc E_r \leftarrow \mc E_r \backslash \{(f_r,b_r)\}$, and $~r \leftarrow r+1$  
\State ~~~~{\bf Else} terminate
% \State {\bf Until} {Some stopping criterion is satisfied}
\State {\bf Output} $\E(\mc E_r)$
\end{algorithmic}
\end{algorithm}
For notational convenience, matrix $\E(\mc E)$ denotes that each element $e_{f,b} = 1,$ when $(f,b) \in \mc E$, otherwise 0. In the worst case, the complexity is approximately $\mc O(B^2F_{\req}^2)$ of the complexity of the inexact BCU-SCA design.
\section{Closed-Form Characterization for Optimal Solution and Analysis}
% {\blue In this section, we first present a closed-form characterization for the optimal solution of the proposed MDS coded design; and then discuss the impact of the system parameters on the optimal delivery policy and network latency; and finally reveal the benefits of MDS codes in contrast with the uncoded design under file splitting.}

\subsection{Closed-Form Characterization for Optimal Solution}
Different from most existing studies on beamformer design in cache-enabled wireless networks,  although the proposed content delivery design $\MDS$ is an MINLP problem, the optimal solution can be characterized by the following Proposition.
\\[+0.1cm]
\emph{
{\bf \emph{Proposition 2:}}  For MDS coded design, let $\big\{\V^*, \E^*, \mb t^*\big\}$ be an optimal solution to $\MDS$. It holds true that 
\begin{align}
      t_f^* &= \frac{\max\limits_b ~ \big\|\|\mb v_{f,b}^*\|_2\big\|_0m'_{f,b}}{ \|\mb s^* \|_1}, ~\forall f \in \F, \label{Nec1}
      % \\
      % {\mathbf{u}}_k^* &= {\mb J_k^{*-1} \mb H_k \mb v_{f_k}^*}\big/ {\big\|\mb J_k^{*-1} \mb H_k \mb v_{f_k}^* \big\|_2}, ~\forall k, \label{Nec2}
\end{align}
where the multicast traffic load vector $\mb s^* = [s_f^*]$ and $s_f^* = \max_b ~ e_{f,b}^* m'_{f,b}$.}
% , and 
% \begin{align}
%    \mb J_k^*  =  {\sum\limits_{f \in {\mathcal{F}_{{\text{req}}}}\backslash \{ {f_k}\} } {{{\mathbf{H}}_k}{{\mathbf{v}}_f^*}{\mathbf{v}}_f^{*H}{\mathbf{H}}_k^H} }  + \sigma _k^2 \mb I .
% \end{align}}

Proposition 2 is a direct deduction of Propositions 1. This Proposition implies that the optimal scheduling of the fronthaul bandwidth allocation, the patterns of SBS collaboration, and the multicast beamformers design are tightly coupled in practice. From the perspective of engineering implementation, this closed-form characterization can help us build the algorithm and reduce computational complexity (e.g., step 4 in Algorithm 1).  
Proposition 2 can also serve as a tool to study the impact of the caching strategy on the optimal delivery policy and network latency. 
% Recall that the traffic load $m_{f,b}' = (1-q_{f,b})S, ~\forall f,b$. Thus, the larger fraction of one file is cached, the fewer fronthaul bandwidth it tends to occupy. 
To provide the insight in greater detail, we analyze two  caching strategies, i.e., the Fractional Cache Distinct (FCD) (see, e.g., \cite{park2016jointJ}), and the Probabilistic Caching (ProbC) (see, e.g., \cite{tao2016content,peng2017layered}), under MDS coding.

{\it Example 1: (FCD Scheme)} This caching scheme aims to boost the cache-hit ratio and provide caching fairness for each requested file. Specifically, each SBS randomly and uniformly stores the same fraction $q$ of each file. That is, a total of $\floor{qn}$ coded packets of file $f$ is randomly cached in SBS $b$.
% i.e., $m_{f,b} = \floor{q n}, \forall f, b$. 
Consider full SBS cooperation, i.e., each user is served by all SBSs simultaneously. By using Proposition 2, we obtain that $1/F_\req$ of the fronthaul bandwidth should be allocated for each file in the fetching of uncached packets. 
Moreover, due to the finite storage in SBSs, $q \leq \mu $ should hold. In this case, the lower bound of the fronthaul latency is $\frac{(1-\mu)SF_\req}{C_F}$.  

{\it Example 2: (ProbC Scheme)} This caching scheme is popularity-aware, i.e.,  each SBS randomly fetches certain fraction of each file by following the file popularity distribution. Specifically, we generate one file index $f'$ by the Zipf distribution at each time; and then randomly cache one coded packet of that file $f'$; and we iterate this process until it reaches the storage limit, i.e., $
\sum_{{f \in \mathcal{F}}} m_{f,b} = \floor{\mu F n},  \forall b \in \cal{B} 
$. Notably, to avoid overlapping and enhance caching efficiency, each SBS can only store unique packets of each file and $m_{f,b} \leq n$ should hold. Clearly, the more popular the file is, the larger fraction of this file is likely to be cached in the SBSs. Recall that the probability of file $f$ being requested is $p_f = cf^{-\gamma}$. Thus, for file $f$, the expected fraction $q_{f,b}$ can be estimated by $ \min \{\frac{\floor{\mu F n}cf^{-\gamma}}{n}, 1\}$. When $\mu Fn$ is an integer, the cache allocation matrix is generally independent of the number of fragments $n$. Moreover, when the popularity order $f$ becomes smaller and the skewness parameter $\gamma$ becomes larger, by using Proposition 2, the required fronthaul bandwidth for delivering file $f$ tends to be narrower.
% Under one special case, where file $f$ is entirely cached by the selected SBSs, i.e., $q_{f,b} = 1$ for any $e_{f,b} = 1$, we naturely have $t_f = 0$. Also, the network latency will be deteriorated if some unpopular files are requested. It is necessary to allocate more fronthaul bandwidth to deliver these unpopular files so as to reduce the fronthaul latency.}  
% =================Uncoded Design===============
% \subsection{Comparison With Uncoded Design}   

% Moreover, the contents transmitted from the CP to each SBS must be in a specific order for file reconstruction. Consequently, the degrees of content reuse is very low, which makes it quite difficult to adopt multicast transmission \cite{liao2017coding,liu2014cache}. Hence, like in \cite{tao2016content,peng2017layered}, unicast transmission is usually adopted in the fronthaul link. {\blue For the fairness of comparison, we also consider the cooperative beamforming, SBS clustering and fronthaul bandwidth allocation in the uncoded design.  
% Accordingly, the latency for uncoded design is given by
% \begin{align}
%     T_{{\text{total}}}^{\text{uncoded}} \left(\U,\V, \mathbf{E}, \mathbf{T}\right) &=   \alpha_E \frac{S}{ \mathop{\min}\limits_{f \in \F}\mathop {\min }\limits_{k \in {{\mathcal{G}}_f}} B_0\phi \left( D_k, J_k \right)}\notag\\ 
%     &+ \alpha_F \mathop{\max }\limits_{f \in {\mathcal{F}_{{\text{req}}}},b \in \mathcal{B}} ~ \frac{e_{f,b} m_{f,b}'}{t_{f, b}C_F + \tau _0}, \label{eq:Ttot_uncoded} 
% \end{align}
%% Single Version
\subsection{Performance Gain Over Fronthaul Unicast}
As aforementioned, many previous content delivery studies investigate cooperative beamforming by adopting separate costly unicast transmissions to fetch the entire contents (e.g., \cite{peng2017layered,tao2016content}) or uncoded fragments (e.g., \cite{park2016joint}). These unicasting transmissions incur higher fronthaul burden which may hinder the collaboration among SBSs from being effective. 
% Besides,  
% e consider the following uncoded design.  
% the contents transmitted from the CP to each SBS must be in a specific order for file reconstruction. Consequently, the degrees of content reuse is very low, which makes it quite difficult to adopt multicast transmission \cite{liao2017coding,liu2014cache}. Hence, like in \cite{tao2016content,peng2017layered}, unicast transmission is usually adopted in the fronthaul link. 
% In this subsection, we elaborate on the performance gain of the proposed MDS aided transmission design over using fronthaul unicast to fetch the uncached contents. 
To investigate the the performance gain of joint design of fronthaul multicast and cooperative beamforming, we consider the following uncoded design where any cache miss is dealt by unicast transmissions.

For the fairness of comparison, we also use fronthaul bandwidth allocation to balance the fronthaul latency for each requested file. Specifically, a fraction $0 \leq t_{f,b} \leq 1$  of fronthaul bandwidth should be allocated for each SBS $b$ to fetch the missing contents of file $f$. We define the fronthaul bandwidth allocation matrix as $\mb T = [t_{f,b}] \in \mathbb{R}^{F_{\req} \times B}$. Moreover, we also consider SBS clustering and multicast beamforming in the edge link. Accordingly, by adopting fronthaul unicast and cooperative beamforming, the network latency for this uncoded design is given by
        \begin{align}
        T_{{\text{total}}}^{\text{FU}} \left(\U,\V, \mathbf{E}, \mathbf{T}\right) & = \alpha_E \frac{S}{ \mathop{\min}\limits_{f \in \F}\mathop {\min }\limits_{k \in {{\mathcal{G}}_f}} B_0\phi \left( D_k, J_k \right)} 
        \notag\\&+ \alpha_F \mathop{\max }\limits_{f \in {\mathcal{F}_{{\text{req}}}},b \in \mathcal{B}} ~ \frac{e_{f,b} m_{f,b}'}{t_{f, b}C_F + \tau _0}. \label{eq:Ttot_uncoded} 
    \end{align}
As can be observed, the utilization of fronthaul unicast may lead to a low spectrum efficiency in the fronthaul links in contrast with the latency \eqref{eq:Ttot} in MDS coded design (i.e., using the same spectrum $t_fC_F$ to send uncached contents). 
% To take advantage of BS cooperation and multicast beamforming in the edge link, 
% Consequently, 
Hence, the optimization problem of uncoded design is given as
\begin{subequations}
  \label{MinT_uncoded} 
  \begin{align}
    \un:~& \mathop {\min }\limits_{\U,\V,\mathbf{E},\mathbf{T}} ~~~~T_{{\text{total}}}^{\text{FU}} \left(\U,\V, \mathbf{E}, \mathbf{T}\right)\\
    &\text{s.t.}~~
    \sum_{f \in {\mathcal{F}_{{\text{req}}}}, b \in \B} t_{f,b}  = 1, 0 \leq t_{f,b} \leq 1,  \forall f, b \label{Un_b}, \\
    & ~~~~~~\eqref{P_b}-\eqref{P_f}.
    \end{align}
\end{subequations}
Since problem $\un$ takes the similar mathematical form to problem $\MDS$, it can also be handled by using the idea of the proposed inexact BCU-SCA design (see Algorithm 1). 
% {\red revised:(Unlike in MDS coded design $\cP_0$, no extra challenge is introduced for the solution of the problem \eqref{MinT_uncoded}.)} 
Moreover, by the principles of Propositions 1 and 2, the following proposition can also be easily obtained. 
\\[+0.1cm]
{\bf \emph{Proposition 3:}}  \emph{Let $\big\{\V^*, \E^*, \mb T^*\big\}$ be an optimal solution to problem $\un$. It holds true that }
\begin{align}
      t_{f,b}^* &= \frac{ ~ \big\|\|\mb v_{f,b}^*\|_2\big\|_0m'_{f,b}}{ \|\vecc(\mb S^*) \|_1}, ~\forall f \in \F, b \in \B, \label{Nec3}
      % {\mathbf{u}}_k^* &= \frac{J_k^{*-1} \mb H_k \mb v_{f_k}^*} {\|J_k^{*-1} \mb H_k \mb v_{f_k}^* \|_2}, ~\forall k, 
\end{align}
\emph{where the load matrix $\mb S^* = [s_{f,b}^*]$, and $s_{f,b}^* = e_{f,b}^* m'_{f,b}$.} 
% \\[+0.1cm]
\vspace*{0.1cm}
% Besides, each optimal receive combiner ${\mathbf{u}}_k^*$ is obtained by \eqref{Nec2}. }

% \begin{align}
%    J_k  =  {\mathbf{u}}_k^{*H}\bigg( {\sum\limits_{f \in {\mathcal{F}_{{\text{req}}}}\backslash \{ {f_k}\} } {{{\mathbf{H}}_k}{{\mathbf{v}}_f^*}{\mathbf{v}}_f^{*H}{\mathbf{H}}_k^H} } \bigg){{\mathbf{u}}_k^*} + \sigma _k^2.
% \end{align}
% Through Propositions 4 and 5, the couping of fronthaul bandwidth allocation, dynamic SBS clustering, and multicast beamforming can be straightforward validated. 
% {\bf \emph{Remark 3:}}  By comparing \eqref{Nec3} with \eqref{Nec1}, it shows that the employment of MDS codes can not only reduce the traffic load, but also improve the spectrum efficiency, leading to a lower latency for content delivery.
According to Propositions 2 and 3, we have the following Corollary.\\[+0.1cm]
{\bf \emph{Corollary 1:}} \emph{Given that a fixed SBS clustering matrix $\E^*$ and capacity-limited fronthaul, the proposed MDS coded design $\MDS$, i.e., multicast the missing packets to SBSs for file reconstruction, 
can achieve a gain of ${\|\vecc(\mb S^*)\|_1}/{\|\mb s^*\|_1}$ for fronthaul latency, in contrast to such an uncoded design $\un$, i.e., send the uncoded fragments to SBSs individually for file reconstruction. In particular, for full BS cooperation and homogeneous  caching, i.e., using FCD scheme and $m_{f,b} = m, \forall f,b$, the fronthaul latency arising from fetching the MDS coded packets can be only as much as $1/B$ of that of fetching the uncoded fragments.}

Moreover, we also analyze another case, where popular files are entirely cached in SBSs, i.e., $m_{f,b} \in \{0,n\}$. When the uncached files are handled by costly fronthaul unicast (e.g., \cite{tao2016content,peng2017layered}), Proposition 3 shows that the fronthaul latency is $n_1S/C_F$, where $n_1$ is the total number of files that all SBSs need to fetch from the CP and $0 \leq n_1\leq BF_{\req}$ holds. Consider the multicast scenario, where the uncached files are fetched by multicast transmissions via fronthaul; by using Proposition 2, the fronthaul latency is $n_2S/C_F$, where $n_2$ is the number of files that should be fetched from the CP and $0 \leq n_2\leq F_\req$ hold. 
Notably, different patterns of SBS collaboration, i.e., $\mb E^*$, may also give rise to different $n_1$ and $n_2$. 
% \vspace*{0.1cm}
% This Corollary 1 implies that under the same SBSs cooperative patterns, the utilization of fronthaul multicast, enabled by employing MDS codes, can help to reduce the fronthaul cost compared with unicast transmission of uncoded fragments.  
% but also give insights into the performance gain of employing MDS codes over uncoded fragments in physical-layer transmissions.

\section{Performance Evaluations}
In this section, we provide numerical simulations to examine the performance of the proposed designs. Similar to \cite{hu2017joint}, we consider a network covering a square area $[-1 \text{km}, 1 \text{km}] \times [-1 \text{km}, 1 \text{km}]$, where $B$ SBSs and $K$ users are randomly and uniformly placed in this area. 
Furthermore, no user is present in each SBS within a radius of 10 m. 
The wireless  channel between SBS $b$ and user $k$ is modeled as
% \begin{align}
%   \mb h_{k,b} = 10^{-L(d_{k,b})/20} \sqrt{\kappa \chi }\mb g_{k,b},
% \end{align}
  $ 
  \mb H_{k,b} = 10^{-L(d_{k,b})/20} \sqrt{\kappa \chi }\mb G_{k,b},
  $
where $L(d_{k,b})$ denotes the path-loss w.r.t. distance $d_{k,b}$; $\kappa$ is the log-normal shadowing parameter; $\chi$ is the antenna power gain, and $\mb G_{k,b}$ is the small-scale fading coefficient. 
% The main parameters for the cached cloud RAN are shown in Table I. 
In our simulation, we consider the following default parameters: for each user, $L(d_{k,b}) = $ 36.8 + 36.7log$(d_{k,b})$; $\kappa = 7$ dB and $\chi = 5$ dBi; $g_{k,b}$ follows $\mc{CN} (\mb{0, I})$;
bandwidth for edge link $B_0$  is  10 MHz and noise power is  -102 dBm; transmit power in each SBS $P_b$ is limited by 1 W; the total number of files in library is 100; all SBSs are assumed to be equipped with the same fractional caching capacity $\mu$. Users' requests from the library follow the Zipf distribution. 
% {\blue For the ease of implementation, we consider
% that each file is split into the same number of fragments, i.e., $n = 5, \forall f \in \cal{F}$, and coded into $m_f$ packets with size of $100$ MB, which is specified by \eqref{MDS}. For the FCD scheme, each SBS $b$ stores $\floor{\mu n}$ unique packets of each file randomly and without any replacement, i.e., $m_{f,b} = \floor{\mu n}, \forall f, b$. For the ProbC scheme, each SBS sequentially fetches unique packets of a certain file based on the Zipf distribution, until it reaches the storage limit. Note that each SBS can only cache $n_f$ unique packets of each file at most, i.e., $0 \leq m_{f,b} \leq n_f$. 
% As a result,
% $
% \sum_{{f \in \mathcal{F}}} m_{f,b} = \mu F n,  \forall b \in \cal{B} 
% $.}
For content placement phase, we adopt the FCD and ProbC strategies. The detailed implementation of each caching strategy could be referred to Sec. V. Without loss of generality, we consider that each file is split into the same number of fragments, i.e., $n = 5, \forall f \in \cal{F}$, and coded into $m_f$ packets with size of $100$ MB, which is specified by \eqref{MDS}. Moreover, from the perspective of mobile users, fronthaul latency and edge latency are considered to have the same importance in our experiments, i.e., $\alpha_E = \alpha_F = 1$.
All of the experimental results were obtained by averaging 100 trials. 

\begin{figure}[!h]
  \begin{minipage}{.4\linewidth}
  \centering
    \includegraphics[scale=0.23]{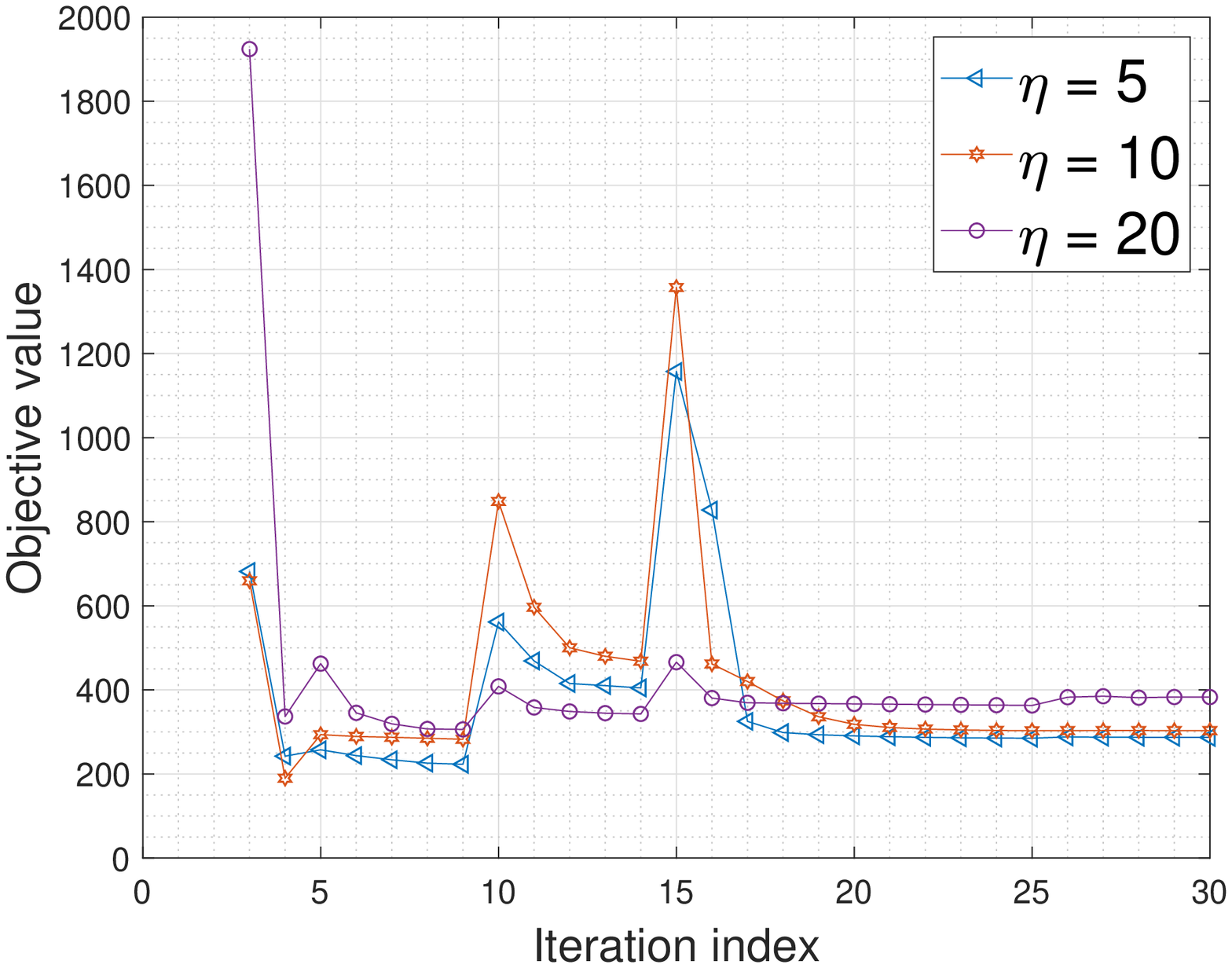}
    \small{(a)}
  \end{minipage}
  ~~~~~
  \begin{minipage}{.5\linewidth}
    \centering
    \includegraphics[scale=0.23]{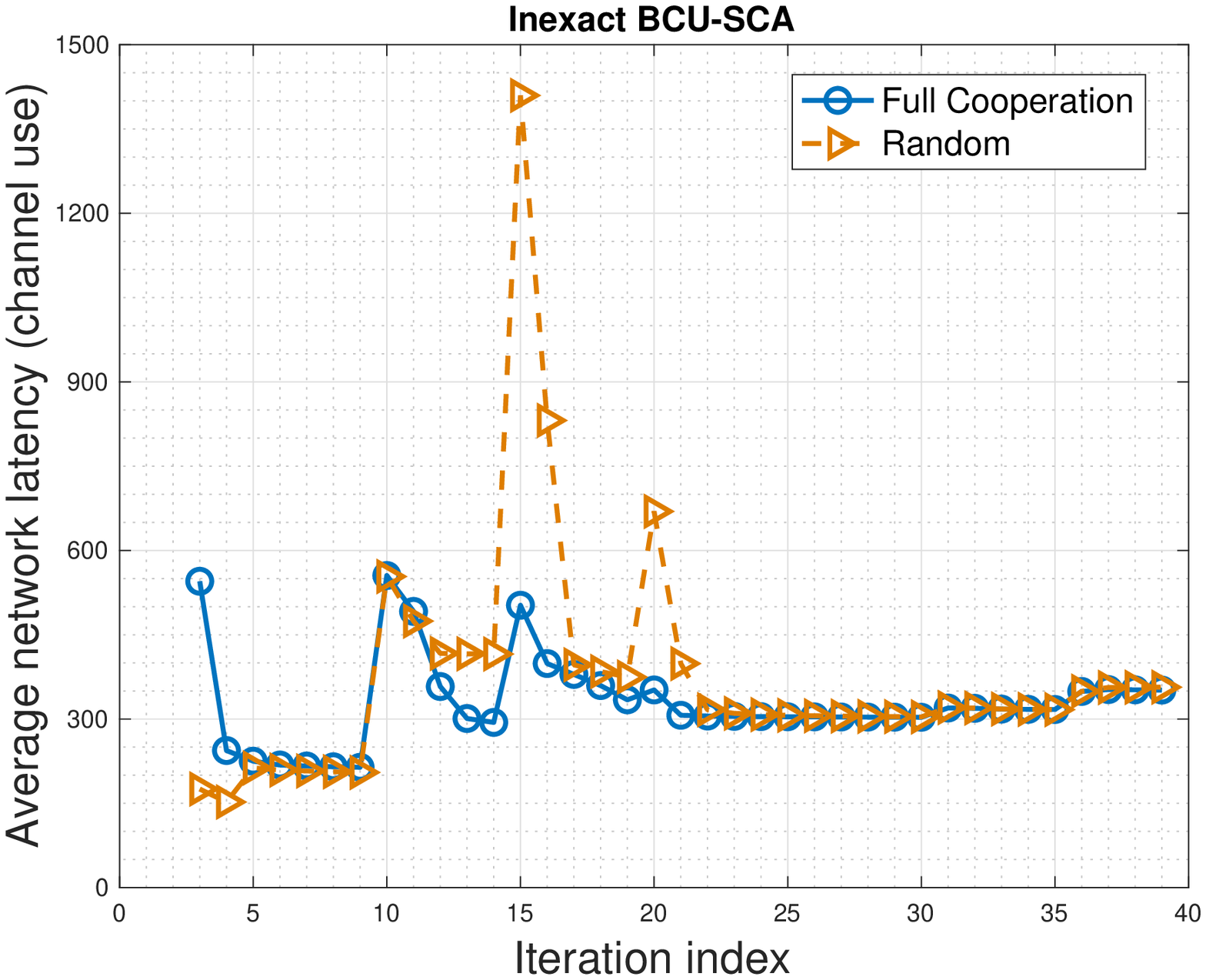}
    \small{(b)}
  \end{minipage}
  \caption{Convergence behavior of inexact BCU-SCA design}\label{Fig:BCUSCA} 
\end{figure}

We first demonstrate the performance of the proposed algorithms for our MDS coded design. 
If the system parameters are not particularly specified, we consider a cache-enabled C-SCNs with the following default scenario: 3 SBSs and 5 active users, which are equipped with $M = 5$ and $N = 3$ antennas, respectively. 
The ProbC is adopted in the considered scenario, where the fractional caching capacity $\mu$ is $20\%$, and content popularity parameter $\gamma = 1$. The capacity of the fronthaul link is set to 10 Mbps. As a comparison, the following algorithms are also implemented to solve problem $\MDS$:
\begin{itemize}
  \item {\bf SB-SCA:} As in \cite{peng2017layered}, binary variable $e_{f,b}$ is replaced by the sparsity of beamforming (SB) vector $\mb w_{f,b}$, and further approximated by the reweighed-$l_1/l_2$ norm. Similar to the inexact BCU-SCA, the SCA technique is adopted to tackle the non-convexity of the rate function. 
  % The entire algorithm runs iteratively, and the final output is projected to the feasible set of $\MDS$.
  \item {\bf NA-SCA:} This algorithm is derived according to \cite{park2017coded,ugur2016cloud}, in which a joint scheme is investigated under given user association strategies, i.e., each $e_{f,b}$ is prefixed. Here, we consider the nearest association (NA), where users access their nearest SBSs, regardless of the local caches. By applying the SCA technique, NA-SCA is used for the evaluation of the upper bound of problem $\MDS$.    
\end{itemize}
Firstly, we illustrate the convergence behavior of the inexact BCU-SCA design in Fig. \ref{Fig:BCUSCA}.
In particular, the initial penalty parameter $\lambda = 0.1$ and then is increased by factor $\eta$, after every 5 iterations. 
Herein, we study the impact of factor $\eta$ on the algorithm. Each $e_{f,b}$ is given by random initialization, i.e. uniformly generated within $[0,1]$, but remains fixed for each trial. As depicted in Fig. \ref{Fig:BCUSCA}(a), the objective value in each trial first monotonically decreases within five iterations (for fixed $\lambda$), and is then lifted due to the increase in the penalty parameter $\lambda$. Furthermore, a larger $\eta$  may lead to a faster convergence speed but is more likely to give a suboptimal solution; while a smaller one could give a more accurate solution but slow down the convergence speed. 
To obtain a better result, we empirically set $\eta = 5$ in all other cases.
We also compare the results between full cooperation initialization, i.e., each $e_{f,b} = 1$, and random initialization. As can be seen in Fig. \ref{Fig:BCUSCA}(b), each curve converges to the same value regardless of the initialization. This finding indicates that the proposed design is not sensitive to the initial setting.
% \begin{figure}[h]
%      \centering
%     \includegraphics[scale=0.4]{Figure/Journal/SCA.eps}
%     % [width=0.3\linewidth, height=0.15\textheight]{prob1_6_2}
%     \caption{Convergence behavior of inexact BCU-SCA design}
%     \label{Fig:BCUSCA} 
% \end{figure}
% \begin{figure}[h]
%         \centering
%         \includegraphics[width=7cm,height=5.9cm]{Figure/Journal/BS.eps}
%         % [width=0.3\linewidth, height=0.15\textheight]{prob1_6_2}
%         \caption{Average network latency versus the number of SBSs.}
%         \label{Fig:BSs}
% \end{figure}

\begin{figure}[!h]
        \centering
        \includegraphics[width=6.8cm,height=5.4cm]{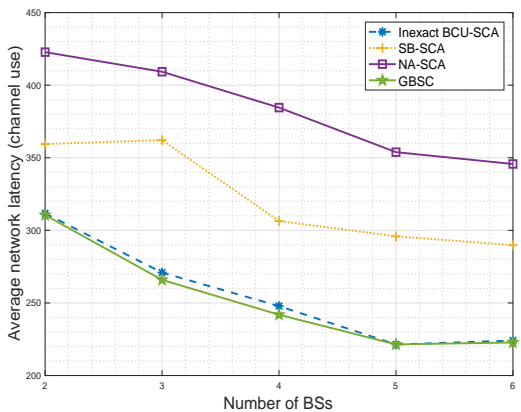}
        \caption{Average network latency versus the number of BSs.}
        \label{Fig:BSs}
\end{figure}

In Fig. \ref{Fig:BSs}, we compare the performance of different algorithms with the number of SBSs. It can be seen that the latencies achieved by the two proposed algorithms are lower than those of the existing SB-SCA and NA-SCA. This observation demonstrates the effectiveness of the proposed variational reformulation of binary constraints and the penalty-based methods. 
% In particular, the proposed algorithms achieve 91 -- 143 less channel uses than NA-SCA and 12 -- 91 less channel uses than SB-SCA.
When the number of SBSs increases, the average latency reduces consistently. This is because more SBSs produce more opportunities for BS cooperation and higher spatial degrees of freedom. 
% Notably, the size of the gap between the two proposed algorithms increases, because some high-rank solutions may be achieved for the inexact BCU-CCCP design  as the dimensions of the variables grow, thereby resulting in a degradation of performance of the SDR. 
More importantly, compared with the GBSC design, the inexact BCU-SCA design achieves almost comparable results with significantly lower complexity. 

\begin{figure*}[t!]
\begin{minipage}{.245\linewidth}
          \centering
          \includegraphics[scale=0.235]{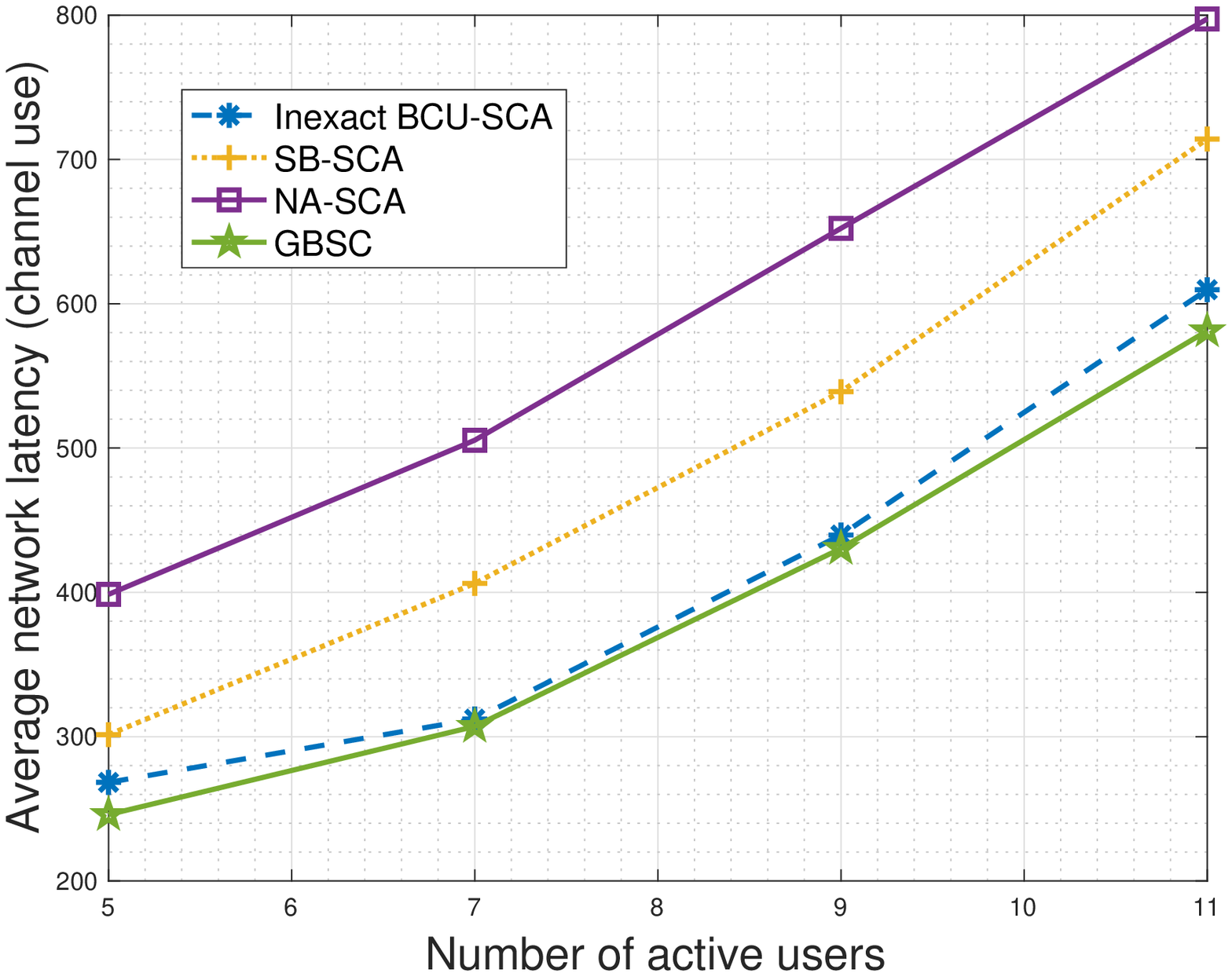}
          % [width=0.3\linewidth, height=0.15\textheight]{prob1_6_2}
          \caption{Effects of the number of active users.}
          \label{Fig:Users}
\end{minipage}
\begin{minipage}{.245\linewidth}
      \centering
      \includegraphics[scale=0.2295]{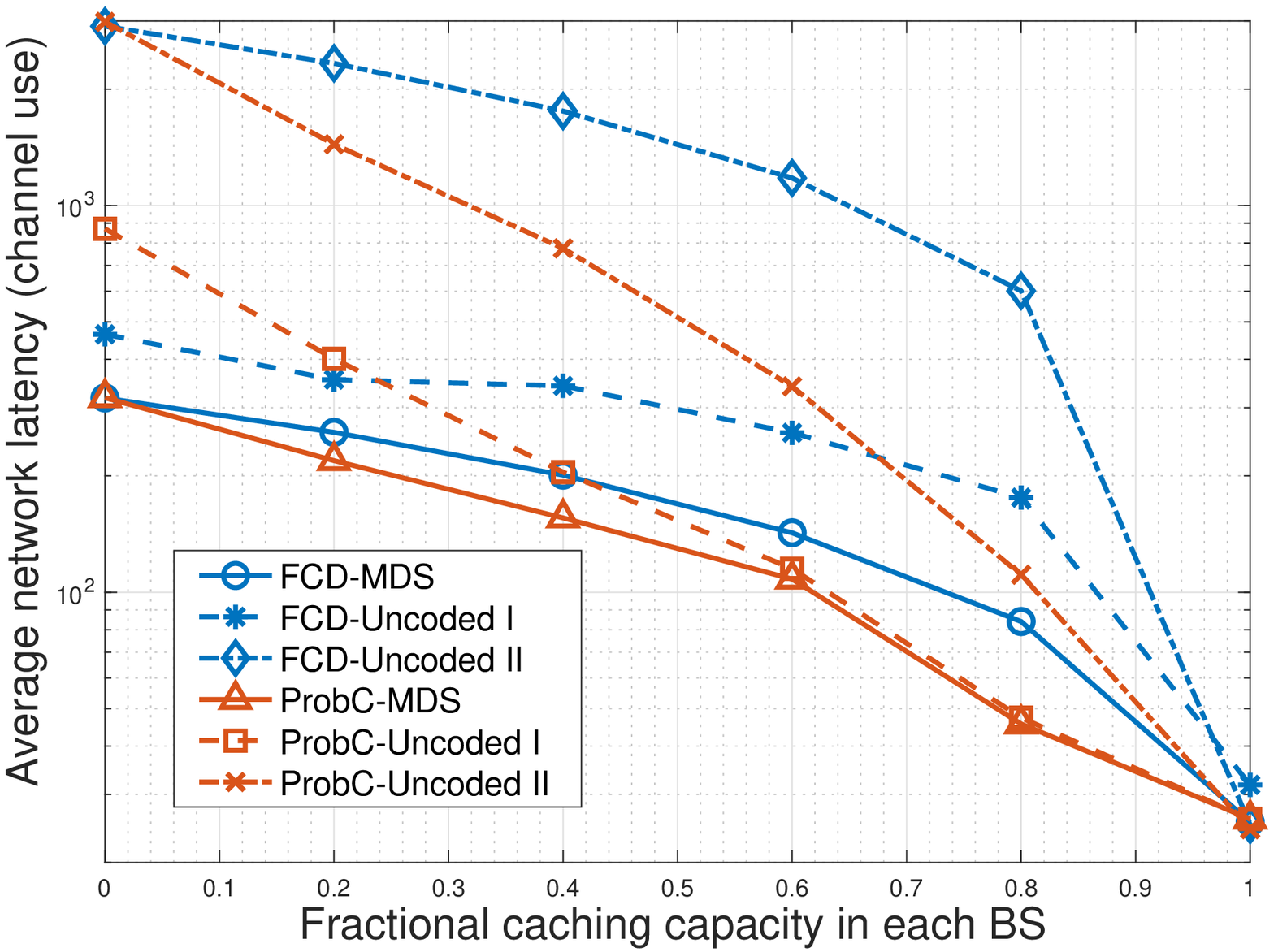}
      % [width=0.3\linewidth, height=0.15\te{}xtheight]{prob1_6_2}
      \caption{Effects of the caching capacity of each BS. } 
      \label{Fig:Mu}
\end{minipage}
\begin{minipage}{.246\linewidth}
      \centering
    \includegraphics[scale=0.23]{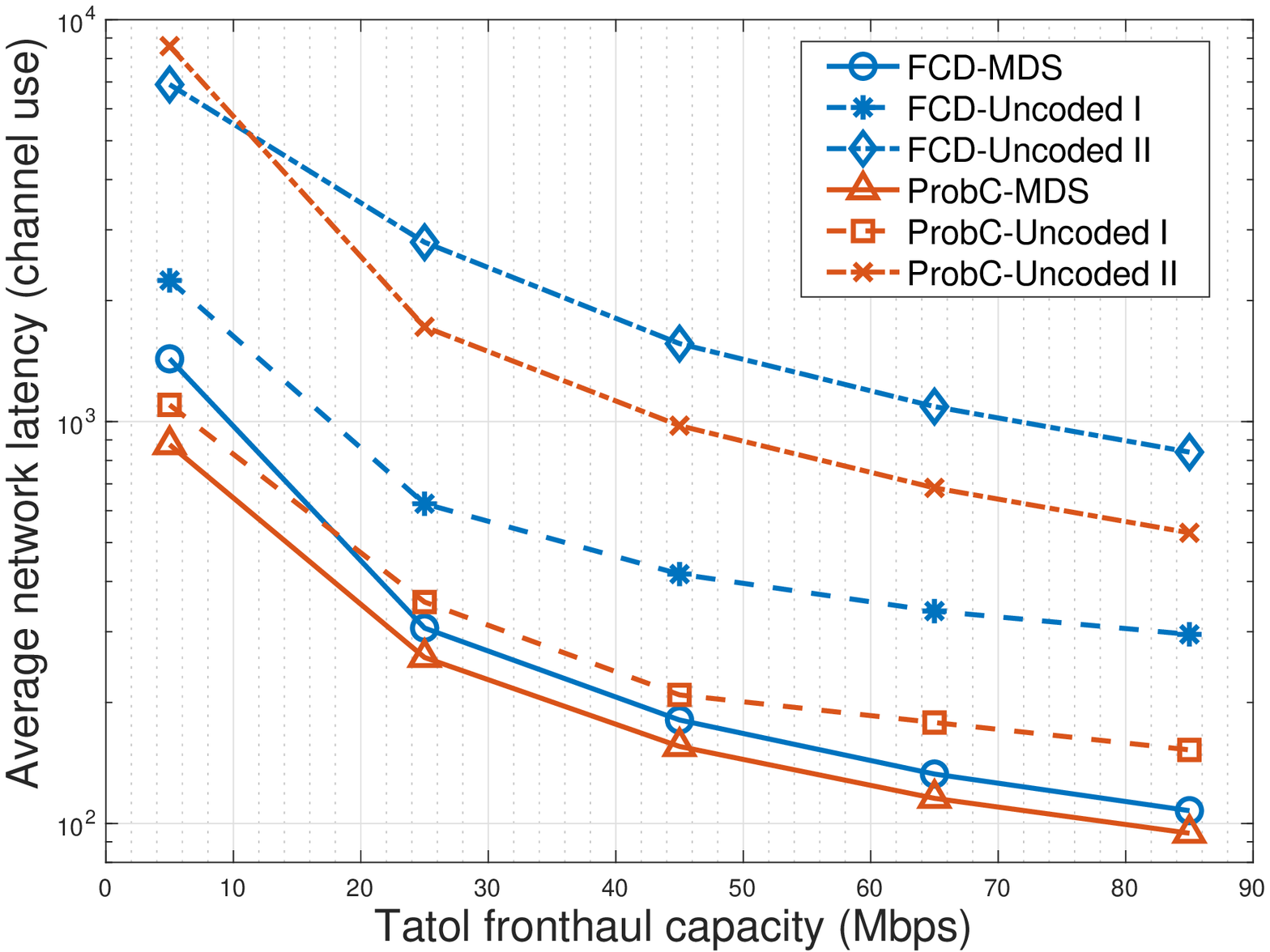}
    % [width=0.3\linewidth, height=0.15\textheight]{prob1_6_2}
    \caption{Effects of the total fronthaul capacity.}
    \label{Fig:FHC}
\end{minipage}
\begin{minipage}{.245\linewidth}
    \centering
    \includegraphics[scale=0.23]{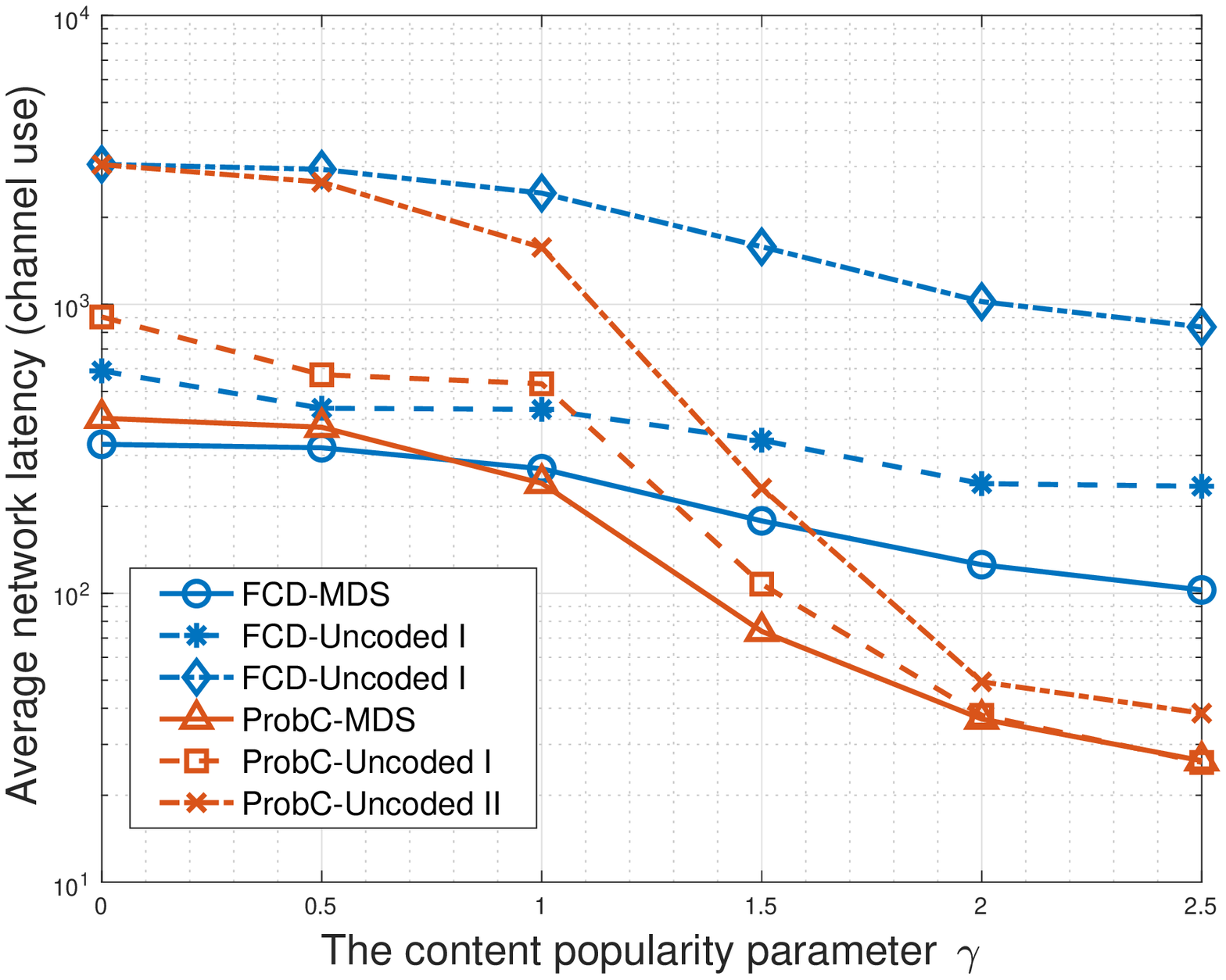}
    % [width=0.3\linewidth, height=0.15\textheight]{prob1_6_2}
    \caption{Effects of file popularity parameter. }
    \label{Fig:Gamma}
\end{minipage}
\end{figure*}

In Fig. \ref{Fig:Users}, we demonstrate the performance of the proposed algorithms with different numbers of active users. The proposed algorithms are observed to achieve lower latency compared with SB-SCA and NA-SCA again. In particular, when there are more users making requests, the size of the gaps between the two proposed designs and SB-SCA increases. Therefore, this implies that the proposed algorithms provide a greater advantage over existing algorithms in a large system. 
% Note that the inexact BCU-CCCP design achieves comparable results to the other proposed one in this scenario. The possible reason is that the dimension of each $\{\mb W_f\}$ is independent of the number of active users. 
Moreover, the inexact BCU-SCA design also obtains results that are very close to that of the GBSC, which confirms the effectiveness of the penalty-based method.   

{Based on the above comparisons, we conclude that the proposed algorithms exhibit superior performance over existing schemes. 
% Moreover, the inexact BCU-SCA design can provide superior results compared with the other proposed algorithm in both performance and computation complexity. 
Note that the inexact BCU-SCA design can provide comparable results with significant lower complexity in contrast to GBSC design. 
Hence, in the following simulation, we only consider adopting the inexact BCU-SCA design.}

% \begin{figure}[h]
%   \begin{minipage}{.5\linewidth}
%           \centering
%           \includegraphics[scale=0.4]{Figure/Journal/Users.eps}
%           % [width=0.3\linewidth, height=0.15\textheight]{prob1_6_2}
%           \caption{Average network latency versus the number of active users.}
%           \label{Fig:Users}
%   \end{minipage}
%   \\
%   \begin{minipage}{.5\linewidth}
%       \centering
%       \includegraphics[scale=0.41]{Figure/Journal/Mu_journal.eps}
%       % [width=0.3\linewidth, height=0.15\te{}xtheight]{prob1_6_2}
%       \caption{Trade-off between the average network latency and the caching capacity of each BS. } 
%       \label{Fig:Mu}
%   \end{minipage}
% \end{figure}

% \subsection{Performance Comparison of the Proposed Delivery Design}
% \subsection{Performance Comparison With Uncoded Designs}

In particular, we consider a large system with 10 SBSs an 15 active users, which in turn, have $M$ of 8 and $N$ of 3. Furthermore, we set the default capacity of the fronthaul link to 30 Mbps, the fractional capacity to $20\%$ and the content popularity parameter as 1, if not specified otherwise. 
% {\blue Under different caching strategies, the proposed MDS coded design is compared with the following two uncoded designs, where any cache miss is dealt by adopting separate costly unicast transmissions:} {\blue To evaluate the benefits of fronthaul multicast and cooperative transmissions,} 
To evaluate the benefits of the proposed MDS coded transmission scheme over the alternative case where any uncached fragments are dealt by adopting separate costly unicast transmissions, we consider the following uncoded designs:
\begin{itemize}
  \item {Uncoded Design I (Uncoded I)}: This design can be completed by solving $\un$. In the edge link, we also consider SBS clustering and multicast beamforming. Different from the MDS coded scheme, the CP adopts unicast to deliver uncoded fragments in the fronthaul link. This design aims at evaluating the potentials of performing cooperative beamforming under fronthaul multicast. The reader is referred to Sec. V for greater detail.
  \item {Uncoded Design II (Uncoded II)}: 
  % Similar to Uncoded I, the CP delivers the associated fragments by unicasting.
  To further evaluate the potentials of effective SBS clustering, all SBSs are allowed to fully cooperate. 
  % This scheme can be completed by solving $\un$, with fixed $e_{f,b} = 1$, for $\forall f, b$.
  Accordingly, we solve $\un$, with fixed $e_{f,b} = 1$, for $\forall f, b$. 
\end{itemize}
% \begin{figure}[h]
%   \begin{minipage}{.5\linewidth}
%       \centering
%     \includegraphics[scale=0.4]{Figure/Journal/FHC_Journal.eps}
%     % [width=0.3\linewidth, height=0.15\textheight]{prob1_6_2}
%     % \caption{Average network latency versus the total fronthaul capacity.}
%     \caption{Effects of the total fronthaul capacity.}
%     \label{Fig:FHC}
%   \end{minipage}
%   \\
%   \begin{minipage}{.5\linewidth}
%     \centering
%     \includegraphics[scale=0.4]{Figure/Journal/Gamma_Journal.eps}
%     % [width=0.3\linewidth, height=0.15\textheight]{prob1_6_2}
%     % \caption{Average network latency versus content popularity parameter. }
%     \caption{Effects of file popularity parameter. }
%     \label{Fig:Gamma}
%   \end{minipage}
% \end{figure}

% Double
\subsubsection{Impact of Cache Storage}
In Fig. \ref{Fig:Mu}, we compare the proposed designs with the uncoded designs under different cache storages. Under both caching strategies FCD and ProbC, the MDS coded designs achieve lower latency than the other uncoded designs, especially in the low cache storage region. In particular, in the absence of caching capability at SBSs, i.e., $\mu = 0$, the MDS coded design obtains only approximately 10.66$\%$ latency of the uncoded design II, under two caching strategies. 
This observation indicates the superior benefits of fronthaul multicast and SBS clustering. In contrast to the FCD, the ProbC is popularity-aware. Hence, when $\mu \geq  0.2$, it 
tends to provide rich opportunities for storing the requested contents, and thus achieves low latency. Moreover, uncoded design II obtains the poorest result in general, because all SBSs are enabled for cooperative transmissions without carefully clustering, and this situation may lead to a significant increase in the fronthaul load. 
Note that when the cache storage increases, the advantages of the proposed designs tend to fade gradually, because the number of requested contents that can be accessed from the SBSs increases without experiencing fronthaul latency. 

\subsubsection{Impact of Fronthaul Capacity} Fig. \ref{Fig:FHC} demonstrates the effects of the fronthaul capacity under different schemes. The proposed design outperforms uncoded designs in the entire curves under each caching strategy. 
% When the fronthaul capacity is 5 -- 25 Mbps, the fronthaul link might be saturated, and thus fronthaul latency constitutes the bottleneck in content delivery. 
Specifically, when the fronthaul capacity is 5 Mbps, the latency achieved by the ProbC-MDS is reduced by 89.91$\%$ compared with the ProbC-Uncoded II. This demonstrates that the fronthaul multicast improves bandwidth efficiency compared with the unicast transmission. Moreover, the notable gap between ProbC-Uncoded I and ProbC-Uncoded II confirms the effectiveness of SBS clustering, especially at limited fronthaul capacity. Moreover, when the fronthaul capacity is larger than 45 Mbps, no significant decrease in latency is observed in each scheme. Hence,  edge latency starts to dominate content delivery. 
% Considering the expensive cost of upgrading the fronthaul link, mobile operators can explore a trade-off between cost and system performance from these simulation results.

 % the ProbC-MDS obtains only 79.64 $\%$ and 10.19 $\%$ latency of ProbC-Uncoded I and ProbC-Uncoded II, respectively. 
% \begin{figure}[h]
%   \centering
%   \includegraphics[scale=0.4]{Figure/Journal/Gamma_Journal.eps}
%   % [width=0.3\linewidth, height=0.15\textheight]{prob1_6_2}
%   \caption{Average network latency versus content popularity parameter. }
%   \label{Fig:Gamma}
% \end{figure}
\subsubsection{Impact of Content Popularity} Fig. \ref{Fig:Gamma} illustrates the performance of all schemes under the variation of the popularity parameter $\gamma$.
% it shows that the proposed design outperforms uncoded designs in the entire curves for both two caching strategies. 
% the performance of all schemes under the variation of popularity parameter $\gamma$.
Clearly, when the content popularity parameter $\gamma$ increases, the average latency reduces gradually, because users make requests following an increasingly centered popularity distribution, and thus few multicast groups are formed. 
% Subsequently, the spectrum efficiency will be improved, which gives a lower latency in the edge link.
Hence, edge latency is substantially reduced. 
Interestingly, when $\gamma \geq 1$, the performance of the ProbC strategy degrades dramatically, while the FCD strategy only witnesses a relatively flat decrease. The reason is for this that the requested contents from users are available in the SBSs with higher opportunity compared with the FCD under the ProbC strategy. Obviously, this fact gives a reduction for both fronthaul and edge latency, which makes curves degrade faster than FCD. When $\gamma \geq 2$, the MDS coded design performs nearly the same as uncoded design I, but still better than uncoded design II. This finding indicates that all requests are almost available in SBSs already, and total latency is dominated by edge transmission. Therefore, the gap between ProbC-MDS and ProbC-Uncoded II reveals the potentials of proper SBS clustering. 
\section{Conclusion}
% In this paper, we develop a unified optimization framework to achieve low latency for content delivery in cached cloud RAN, by utilizing MDS codes and cooperative transmissions. We formulate the problem to minimize latency of both fronthaul and edge links, which takes fronthaul multicast, SBS clustering, and beamforming into account. To solve the resulting MINLP, two penalty-based algorithms are derived with the polynomial complexity. Furthermore, we demonstrate the necessary conditions for the optimal solution, which can quantify the benefits of MDS coded design. Simulation results show the superior performance of the proposed designs.  

In this paper, we have developed an MDS codes-aided transmission scheme in cache-enabled C-SCNs, by exploiting fronthaul multicast and cooperative beamforming. We have formulated the problem to minimize latency of both fronthaul and edge links under physical layer transmission and fronthaul bandwidth constraints. To solve the resulting mixed-integer nonlinear program, a penalty-based design has been derived with low complexity and convergence guarantee. Building upon the solution of the inexact BCU-SCA design, the greedy SBS clustering design has been further developed to improve the solution of BCU-SCA, yet at the cost of computational complexity.
Finally, closed-form characterization of the optimal solution has been investigated, from which the performance gain of MDS codes has been corroborated. Simulation results have been presented to demonstrate the superior performance of the proposed designs.
 
Moreover, it is worth pointing out that the derived closed-form characterization of the optimal solution reveals that the caching strategy impacts the collaborative patterns of SBSs, beamformer design, and fronthaul bandwidth allocation. Indeed, it is promising to study the mixed timescale problem by jointly optimizing cache updating and content delivery so as to further shorten latency and enhance users' quality of service. Nevertheless, to solve such a complex problem, it is very challenging but necessary to develop low-complexity algorithms for engineering implementation in wireless networks. This problem is beyond the scope of this paper and worthy of an independent work. A  preliminary design has been studied in \cite{Wu2019ICC}. In our future work, we shall explore more efficient strategies for the joint consideration of cache updating and content delivery.  

\appendices

\section{Proof for Proposition 1}

% {\bf \emph{Proposition 2}: } Define vector $\mb s = \big\{ \max\limits_b ~\overline e_{f,b} m'_{f,b}\big\}_{f \in \F} \in \mathbb{C}^{F_{\req}}$. 
% Consider constant $\tau_0$ approaches 0 and $\mb s \neq  \mb 0$, thus one optimal solution $(\U^*, \bt^*)$ to problem $\mc{R}_1(\overline{\W}, \overline{\mb{E}})$ can be expressed as
% \begin{align}
%   {\mathbf{u}}_k^* &= {\text{ }}\frac{{\overline {J'} _k^{ - 1}{{\mathbf{z}}_p}\left( {{{\mathbf{H}}_k}{{\overline {\mathbf{W}} }_{{f_k}}}{\mathbf{H}}_k^H} \right)}}{{\left\| {\overline {J'} _k^{ - 1}{{\mathbf{z}}_p}\left( {{{\mathbf{H}}_k}{{\overline {\mathbf{W}} }_{{f_k}}}{\mathbf{H}}_k^H} \right)} \right\|_2}} \label{OP:u}\\
%   t_f^* &= 
%   \frac{\max\limits_b ~\overline e_{f,b} m'_{f,b}}{ \|\mb s\|_2} \label{OP:t}
% \end{align}
% where operator $\mb z_p (\mb X)$ denotes the principle eignvector of the matrix $\mb X$, and 
% \begin{align}
%   \overline {J'}_k & = \sum\limits_{f \in {\mathcal{F}_{{\text{req}}}}\backslash \{ {f_k}\} }^{} {{\mathbf{u}}_k^H{{\mathbf{H}}_k}{\overline {\mathbf{W}}_f}{\mathbf{H}}_k^H} {{\mathbf{u}}_k} + \sigma _k^2,
% \end{align}
% for $ \forall k \in \K$.

Note that problem $\R_3$ can be decomposed into three subproblems, which are given as follows
% \begin{subequations}
%   \begin{align}
%     \mc{R}_2 (&\overline{\W}, \overline{\mb{E}}): \mathop {\min }\limits_{ \mb{U}, \mb{t}, t_E, t_F, \mb{r}}~~ \alpha_E t_E + \alpha_F t_F \\
%     \text{s.t.}   ~~& {t_F} \geq \frac{{{{\overline e }_{f,b}}m_{f,b}'}}{{{t_{f}}{C_F} + {\tau _0}}}, \forall f,b\\
%     & g_k^1 (\U, \overline{\W}, \br)  - g_k^2 (\U, \overline{\W})  \leq 0, ~~\forall k,\\ 
%     & \eqref{P_c}, \eqref{P_g}, \eqref{P_01e}, 
%     \end{align}\label{Relax_2}
% \end{subequations}
% \begin{subequations}
\begin{align}
      \mathop {\min }_{ \mc{U}, t_E, \mb{r}}~~ &t_E ~~
    \text{s.t.} ~~
    \{ \eqref{P_c}, \eqref{P_01e}, \eqref{R5_c}\}, \label{B_1}\\
      \min_{\Z}~~  &-2(\vecc(\overline \E) - \mb 1)^T (2\vecc(\Z) - \mb 1) ~~\text{s.t. } ~~ \eqref{RR_d}, \label{BB_3}\\
      \mathop {\min }_{ \mb{t}, t_F}~~ &t_F ~~
    \text{s.t.} ~~
    \{ \eqref{P_g}, \eqref{nR3_b}\}. \label{B_2}
\end{align}
% \end{subequations}

For problem \eqref{B_1}, the optimal solution can be achieved when the rate of each user is maximized. Hence, problem \eqref{B_1} can be solved by optimizing a group of rate maximization problems
% \begin{align}
%     \begin{split}
%     r_k = \mathop {\max }\limits_{ \mb{u}_k}~~ & \log(\overline  D_k + \overline J_k) - \log(\overline J_k) \\
%     \text{s.t.} ~~ &\|\mb u_k\|_2 = 1, 
%     \end{split}
% \label{B_3}
% \end{align}
\begin{align}
    \begin{split}
    r_k = \mathop {\max }\limits_{ \mb{u}_k}~~  \log(\overline  D_k + \overline J_k) - \log(\overline J_k) ~~ \text{s.t.} ~~ \|\mb u_k\|_2 = 1, 
    \end{split}
\label{B_3}
\end{align}
for all $k \in \K$, $\overline  D_k =  {{\mathbf{u}}_k^H{{\mathbf{H}}_k}{\overline {\mathbf{W}}_{{f_k}}}{\mathbf{H}}_k^H{{\mathbf{u}}_k}}$, and $\overline  J_k = \chi_{k,3} (\mb u_k, \overline \W)$.
% \begin{align}
%   \overline  D_k &=  {{\mathbf{u}}_k^H{{\mathbf{H}}_k}{\overline {\mathbf{W}}_{{f_k}}}{\mathbf{H}}_k^H{{\mathbf{u}}_k}},  \\
%   \overline  J_k &= \sum\limits_{f \in {\mathcal{F}_{{\text{req}}}}\backslash \{ {f_k}\} }^{} {{\mathbf{u}}_k^H{{\mathbf{H}}_k}{\overline {\mathbf{W}}_f}{\mathbf{H}}_k^H} {{\mathbf{u}}_k} + \sigma _k^2.
% \end{align}
By now, we observe that an optimal $\mb u^*$ for problem \eqref{B_3} can be obtained by maximizing the signal-to-interference-plus-noise-ratio  (SINR) \cite{gomadam2011distributed}. By the (31) in \cite{gomadam2011distributed}, we have the closed-form expression \eqref{OP:u2}. Then, we elaborate on the deduction for \eqref{OP:z2}. The auxiliary variable $\Z$ is updated by solving problem \eqref{BB_3}, 
where the equality in the constraint of problem \eqref{BB_3} should be active when the optimal solution is achieved. By the Cauchy--Schwarz inequality, we can obtain \eqref{OP:z2}. Regarding problem \eqref{B_2}, it can be rewritten as 
\begin{subequations}
  \label{B4}
  \begin{align}
    \textstyle\min_ {\bt, t_F}~~&  t_F\\
    \text{s.t.   }  ~~ &t_F \geq \frac{\overline s_f}{t_fC_F + \tau_0}, ~\forall f\in \F, \label{B4_b}\\
    &   0 \leq t_f \leq 1, ~\forall f \in \F, \label{B4_c}\\
    &  \textstyle\sum_{f \in \F} t_f = 1, \label{B4_d}
    % & \eqref{P_g}
      \end{align}
\end{subequations}
which is a convex problem. We have the following observations: when $\overline s_f = 0$, it indicates the absence of a fronthaul traffic load for file $f$. Thus, zero bandwidth is assigned for multicast group $f$. For problem \eqref{B4}, as $\tau$ approaches 0, an optimal solution is given by $t_f^* = 0$, when $\overline s_f = 0$. Moreover, when $\overline {\mb s} = \mb 0$, we have the optimal value $t_F^* = 0$, for any feasible solution. 
In the following, we discuss the case with $\overline s_f > 0, \forall f \in \F$. Consider that $\tau$ approaches 0. Hence, $t^*_f>0, ~\forall f \in \F$, otherwise, the positive infinity objective value will be achieved. By constraint \eqref{B4_b}, we have the optimal value $T_F^* > 0$. The Lagrangian function of problem \eqref{B4} is given by
% \begin{align}
%   \mc L &= t_F + \textstyle\sum_{f\in \F}\big[ \mu_f (\frac{\overline s_f } {t_fC_F} - t_F) + \sigma_{f,1}(t_f -1 )- \sigma_{f,2} t_f\big] \notag \\ 
%   &+ \gamma(\textstyle\sum_{f\in\F} t_f -1) \label{LL}
% \end{align}
%% Single version
\begin{align}
  \mc L =& t_F + \textstyle\sum_{f\in \F}\big[ \mu_f (\frac{\overline s_f } {t_fC_F} - t_F) + \sigma_{f,1}(t_f -1 )- \sigma_{f,2} t_f\big] \notag\\&+\gamma(\textstyle\sum_{f\in\F} t_f -1) \label{LL}
\end{align}
where the non-negative Lagrange multipliers $\mu_f,  \sigma_{f,1},  \sigma_{f,2}$ are associated with constraints \eqref{B4_b}, \eqref{B4_c}, respectively; and the Lagrange multiplier $\gamma$ is associated with constraint \eqref{B4_d}.  According to the Karush-Kuhn-Tucker conditions, we differentiate \eqref{LL} w.r.t $t_F$ and $t_f$ and set these partial derivatives to zero, resulting in the following equations
% \begin{align}
%     \sum_{f \in \F} \mu_f &= 1\label{eq:tF}\\
%     - \frac{\overline s_f \mu_f }{t_f^{*2}C_F} + \sigma_{f,1} - &\sigma_{f,2} + \gamma = 0, ~~ \forall f \in \F, \label{eq:tf}
% \end{align}
\begin{align}
    \textstyle \sum_{f \in \F} \mu_f = 1, 
    - \frac{\overline s_f \mu_f }{t_f^{*2}C_F} + \sigma_{f,1} - \sigma_{f,2} + \gamma = 0, \forall f, \label{eq:tf}
\end{align}  
for the optimal solution $\bt^*$ and $t_F^*$. 
By complementary slackness, we also have
% \begin{align}
%   \frac{\overline s_f \mu_f}{t_f^*C_F} = t_F^* \mu_f, ~ \sigma_{f,1} (t_f^* -1) = 0, ~\sigma_{f,2}t_f^* = 0, ~\forall f. \label{eq:cs}
% \end{align}
\begin{align}
  {\overline s_f \mu_f}/{(t_f^*C_F)} = t_F^* \mu_f, ~ \sigma_{f,1} (t_f^* -1) = 0, ~\sigma_{f,2}t_f^* = 0, ~\forall f. \label{eq:cs}
\end{align}
Subsequently, we consider two cases. Case 1: when $ 0< t_f^* < 1$ for all $f \in \F$, it obtains that $\sigma_{f,1} = \sigma_{f,2} = 0$, for all $f\in \F$. Substituting \eqref{eq:cs} into \eqref{eq:tf}, it follows that
\begin{align}
  t_f^* \gamma = t_F^* \mu_f, ~\forall f. \label{eq:tmp}
\end{align}
Recalling \eqref{eq:tf} and \eqref{B4_d}, equation \eqref{eq:tmp} further shows the following results: 
  $\gamma = t_F^*,$ and $t_f^* = \mu_f > 0, ~\forall f.$
As a consequence, based on \eqref{eq:cs}, we obtain
$
  t_F^* = \frac{\overline s_f}{t_f^* C_F}, ~ \forall f \in \F. 
$ Furthermore, by \eqref{B4_d}, we obtain 
$
  t^*_f = \overline s_f/\|\overline {\mb s}\|_1. 
$
Case 2: when there exists one $t_{f'}^* = 1$ for multicast group $f'$, it shows that only file $f'$ is transferred via fronthaul.
By combining all cases discussed above, we conclude that Proposition 1 holds true.  
\bibliographystyle{IEEEtran}
\bibliography{references}
\begin{IEEEbiography}[{\includegraphics[width=1in,height=1.25in,clip,keepaspectratio]{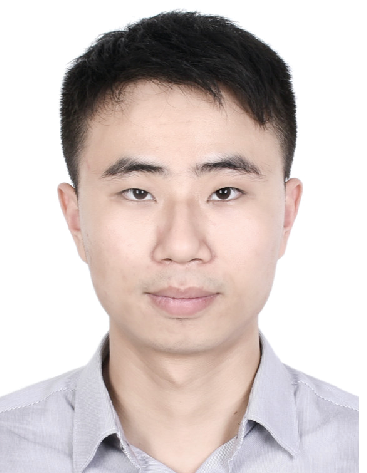}}]{Xiongwei Wu}(S'18)
received the B.Eng. in Electronic Information Engineering from University of Electronic Science and Technology of China (UESTC), Chengdu, China, in 2016. He is currently working toward the Ph.D. degree with the Chinese University of Hong Kong (CUHK), Shatin, Hong Kong SAR, China. From August 2018 to December 2018, he was a Visiting International Research Student with the University of British Columbia (UBC), Vancouver, BC, Canada. He is currently a Visiting Student Research Collaborator with Princeton University, Princeton, NJ, USA. His research interests include signal processing and resource allocation in wireless networks, decentralized optimization, and machine learning.
\end{IEEEbiography}
% \vspace*{-1cm}
\begin{IEEEbiography}[{\includegraphics[width=1in,height=1.25in,clip,keepaspectratio]{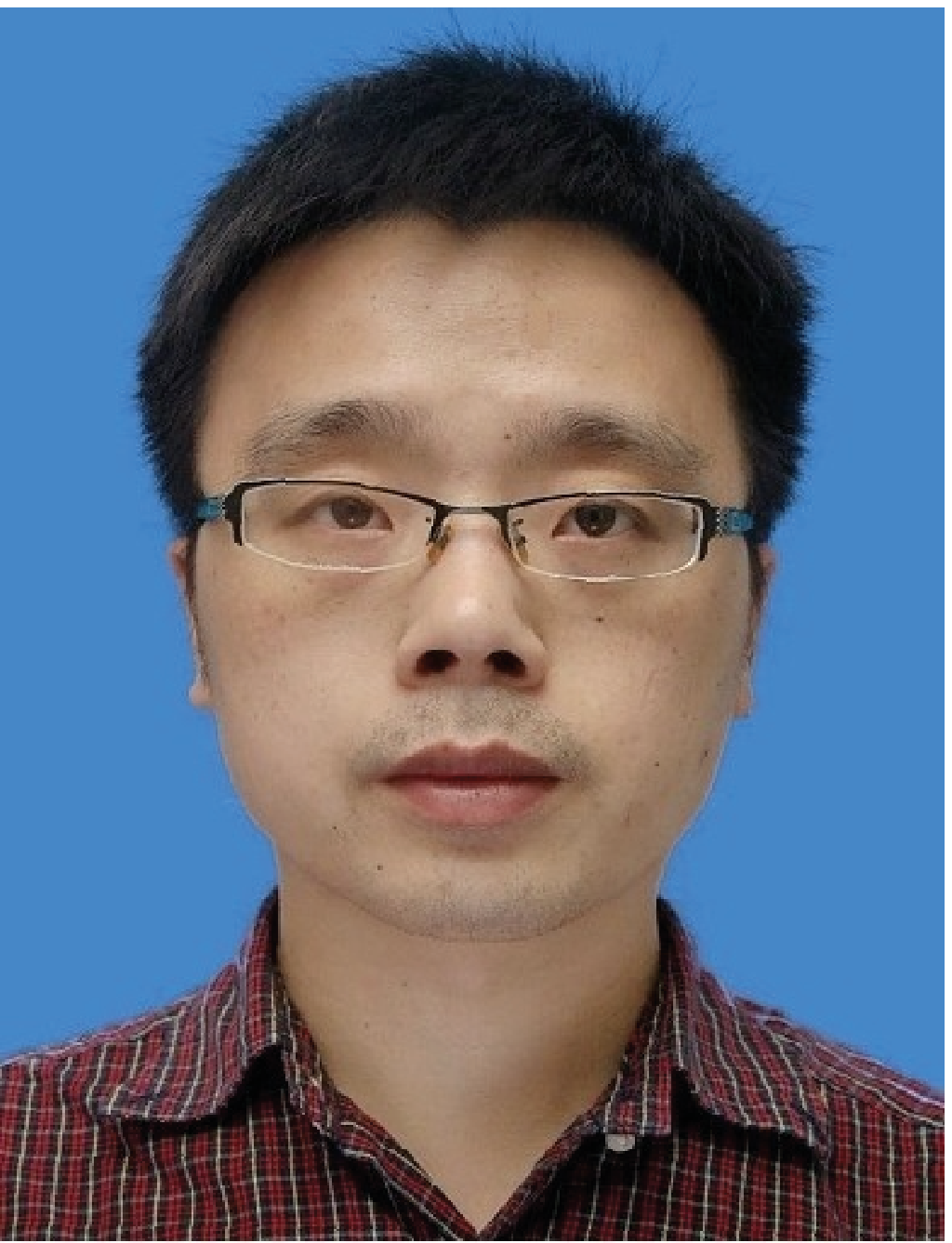}}]{Qiang Li}
received the B.Eng. and M.Phil. degrees in Communication and Information Engineering from University of Electronic Science and Technology of China (UESTC), Chengdu, China, and the Ph.D. degree in Electronic Engineering from the Chinese University of Hong Kong (CUHK), Hong Kong, in 2005, 2008, and 2012, respectively. From August 2011 to January 2012, he was a Visiting Scholar with the University of Minnesota, Minneapolis, MN, USA. From February 2012 to October 2013, he was a Research Associate with the Department of Electronic Engineering and the Department of Systems Engineering and Engineering Management, CUHK. Since November 2013, he has been with the School of Information and Communication Engineering, UESTC, where he is currently an Associate Professor. His research interests include efficient optimization algorithm design for wireless communications and machine learning.

He received the First Prize Paper Award in the IEEE Signal Processing Society Postgraduate Forum Hong Kong Chapter in 2010, a Best Paper Award of IEEE PIMRC 2016, and the Best Paper Award of the IEEE Signal Processing Letters 2016.
\end{IEEEbiography}

\begin{IEEEbiography}[{\includegraphics[width=1in,height=1.25in,clip,keepaspectratio]{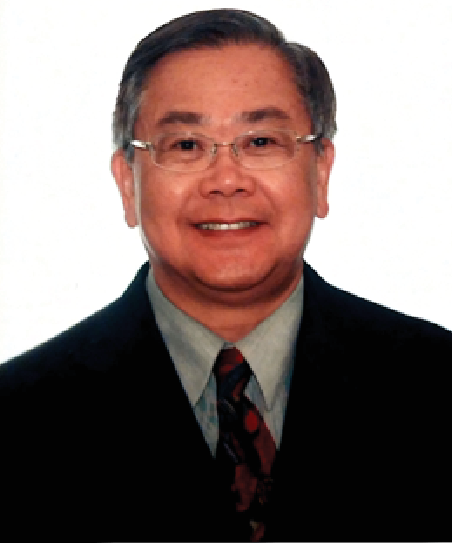}}]{Victor C. M. Leung} [S'75-M'89-SM'97-F'03] received the B.A.Sc. (Hons.) degree in electrical engineering from the University of British Columbia (UBC) in 1977, and was awarded the APEBC Gold Medal as the head of the graduating class in the Faculty of Applied Science. He attended graduate school at UBC on a Canadian Natural Sciences and Engineering Research Council Postgraduate Scholarship and received the Ph.D. degree in electrical engineering in 1982.

From 1981 to 1987, Dr. Leung was a Senior Member of Technical Staff and satellite system specialist at MPR Teltech Ltd., Canada. In 1988, he was a Lecturer in the Department of Electronics at the Chinese University of Hong Kong. He returned to UBC as a faculty member in 1989, and held the positions of Professor and TELUS Mobility Research Chair in Advanced Telecommunications Engineering in the Department of Electrical and Computer Engineering when he retired at the end of 2018 and became a Professor Emeritus. Since Mar. 2019, he has been appointed as a Distinguished Professor in the College of Computer Science and Software Engineering at Shenzhen University, China. Dr. Leung has co-authored more than 1200 journal articles and conference papers, 43 book chapters, and co-edited 14 book titles. Several of his papers had been selected for best paper awards. His research interests are in the broad areas of wireless networks and mobile systems.

Dr. Leung is a registered Professional Engineer in the Province of British Columbia, Canada. He is a Fellow of the Royal Society of Canada, the Engineering Institute of Canada, and the Canadian Academy of Engineering. He was a Distinguished Lecturer of the IEEE Communications Society. He is serving on the editorial boards of the IEEE Transactions on Green Communications and Networking, IEEE Transactions on Cloud Computing, IEEE Access, IEEE Network, Computer Communications, and several other journals, and has previously served on the editorial boards of the IEEE Journal on Selected Areas in Communications - Wireless Communications Series and Series on Green Communications and Networking, IEEE Transactions on Wireless Communications, IEEE Transactions on Vehicular Technology, IEEE Transactions on Computers, IEEE Wireless Communications Letters, and Journal of Communications and Networks. He has guest-edited many journal special issues, and provided leadership to the organizing committees and technical program committees of numerous conferences and workshops. He received the IEEE Vancouver Section Centennial Award, the 2011 UBC Killam Research Prize, the 2017 Canadian Award for Telecommunications Research, the 2018 IEEE TCGCC Distinguished Technical Achievement Recognition Award, and the 2018 ACM MSWiM Reginald Fessenden Award. He co-authored papers that won the 2017 IEEE ComSoc Fred W. Ellersick Prize, the 2017 IEEE Systems Journal Best Paper Award, the 2018 IEEE CSIM Best Journal Paper Award and the 2019 IEEE TCGCC Best Journal Paper Award. He is named in the current Clarivate Analytics list of "Highly Cited Researchers".
\end{IEEEbiography}
\begin{IEEEbiography}[{\includegraphics[width=1in,height=1.25in,clip,keepaspectratio]{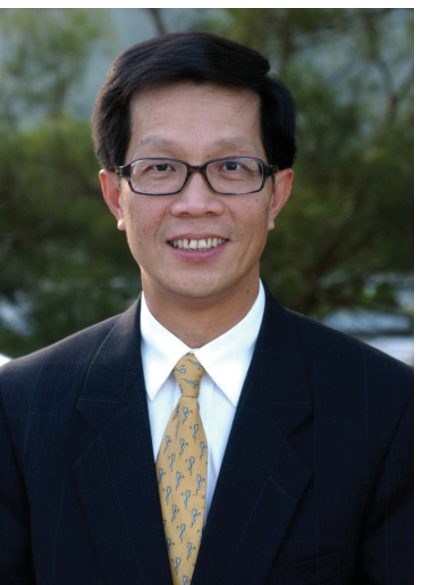}}]{Pak-chung Ching}
received the B. Eng. (1st Class Honors) and Ph.D. degrees from the University of Liverpool, UK, in 1977 and 1981 respectively. From 1981 to 1982, he was Research Officer at the University of Bath, UK. In 1982, Prof. Ching returned to Hong Kong and joined the then Hong Kong Polytechnic as a lecturer. Since 1984, he has been with the Department of Electronic Engineering of the Chinese University of Hong Kong (CUHK), where he is currently Choh-Ming Li Professor of Electronic Engineering. He was Department Chairman from 1995 to 1997, Dean of Engineering from 1997 to 2003 and Head of Shaw College from 2004 to 2008. He became Director of the Shun Hing Institute of Advanced Engineering in 2004. From 2006 till end of 2014, Prof Ching was appointed as Pro-Vice-Chancellor/Vice-President of CUHK. Between 2013 to 2014, Prof. Ching also took up the Directorship of the CUHK Shenzhen Research Institute. Prof. Ching is very active in promoting professional activities, both in Hong Kong and overseas. He was a council member of the Institution of Electrical Engineers (IEE), past chairman of the IEEE Hong Kong Section, an associate editor of the IEEE Transactions on Signal Processing from 1997 to 2000 and IEEE Signal Processing Letters from 2001 to 2003. He was also a member of the Technical Committee of the IEEE Signal Processing Society from 1996 to 2004. He was appointed Editor-in-Chief of the HKIE Transactions between 2001 and 2004. He has been an Honorary Member of the editorial committee for Journal of Data Acquisition and Processing since 2000. Prof. Ching has been instrumental in organizing many international conferences in Hong Kong including the 1997 IEEE International Symposium on Circuits and Systems where he was the Vice-Chairman. He also served as Technical Program Co-Chair of the 2003 and 2016 IEEE International Conference on Acoustics, Speech and Signal Processing. Prof Ching was awarded the IEEE Third Millennium Award in 2000 and the HKIE Hall of Fame in 2010. In addition, Prof. Ching also plays an active role in community services.  He was awarded the Silver Bauhinia Star (SBS) and the Bronze Bauhinia Star (BBS) by the HKSAR Government in 2017 and 2010, respectively, in recognition of his long and distinguished public and community services.  He is presently Chairman of the Board of Directors of the Nano and Advanced Materials Institute Limited, Council Member of the Shaw Prize Foundation, as well as a Member of the Museum Advisory Committee (MAC) and the Chairperson of its Science Sub-committee. He is elected as President of the Hong Kong Academy of Engineering Sciences (HKAES) in 2018. 
\end{IEEEbiography}

% \begin{IEEEbiography}{Michael Shell}
% Biography text here.
% \end{IEEEbiography}

% % if you will not have a photo at all:
% \begin{IEEEbiographynophoto}{John Doe}
% Biography text here.
% \end{IEEEbiographynophoto}

% % insert where needed to balance the two columns on the last page with
% % biographies
% %\newpage

% \begin{IEEEbiographynophoto}{Jane Doe}
% Biography text here.
% \end{IEEEbiographynophoto}

% You can push biographies down or up by placing
% a \vfill before or after them. The appropriate
% use of \vfill depends on what kind of text is
% on the last page and whether or not the columns
% are being equalized.

%\vfill

% Can be used to pull up biographies so that the bottom of the last one
% is flush with the other column.
%\enlargethispage{-5in}

% that's all folks
\end{document}